\documentclass[preprint,times,11pt]{elsarticle}

\usepackage[utf8]{inputenc}
\usepackage[T1]{fontenc}
\usepackage{mathptmx}
\usepackage{microtype}

\usepackage{mathtools}
\usepackage{bm}            
\usepackage{nicefrac}      
\usepackage{siunitx}       
\usepackage{amssymb}
\usepackage[margin=1.0in]{geometry}

\usepackage{graphicx}      
\usepackage[table]{xcolor} 
\usepackage{tikz}

\usepackage{dcolumn}       
\usepackage{tabularx}
\usepackage{booktabs}      
\usepackage{float}         
\usepackage{placeins}      

\usepackage{etoolbox}
\usepackage{printlen}
\usepackage[normalem]{ulem} 
\usepackage{lipsum}         

\usepackage{url}
\usepackage{hyperref}
\usepackage[noabbrev]{cleveref}

\makeatletter
\def\@email#1#2{%
 \endgroup
 \patchcmd{\titleblock@produce}
  {\frontmatter@RRAPformat}
  {\frontmatter@RRAPformat{\produce@RRAP{*#1\href{mailto:#2}{#2}}}\frontmatter@RRAPformat}
  {}{}
}%
\makeatother

\usepackage{subcaption} 

\newcommand{\diff}{\mathrm{d}} 
\newcommand{\pfrac}[2]{\frac{\partial #1}{\partial #2}}
\newcommand{\der}[2]{\frac{\partial #1}{\partial #2}}

\newcommand{\f}[1]{\overline{#1}} 
\newcommand{\fa}[1]{\widetilde{#1}}

\newcommand{\ff}[1]{\widehat{#1}}

\newcommand{\vv}[1]{\boldsymbol{#1}}

\newcommand{\av}[1]{\langle #1 \rangle}
\newcommand{\intl}{\int\displaylimits}
\newcommand{\suml}{\sum\displaylimits}

\newenvironment{rewrite}{
}{
}

\newenvironment{rewriteAdrien}{
}{
}

\newcommand{\rewriteFrancoise}[2]{%
#2
}

\usetikzlibrary{calc}

\definecolor{blue_mpl}  {HTML}{1f77b4}
\definecolor{orange_mpl}{HTML}{ff7f0e}
\definecolor{green_mpl} {HTML}{2ca02c}
\definecolor{red_mpl}   {HTML}{d62728}

\definecolor{hot_mpl} {HTML}{F08080}
\definecolor{cold_mpl}{HTML}{ADD8E6}

\begin{document}
\begin{frontmatter}
	\title{A posteriori study of Thermal-Large Eddy Simulation in solar receiver operating conditions}

	\author[promes,lisn]{Yanis Zatout}
	\author[promes]{Françoise Bataille}
	\author[promes]{Adrien Toutant}
	\corref{Adrien Toutant}

	\begin{abstract}
		This study investigates Thermal-Large Eddy Simulations (T-LES) of anisothermal
		and turbulent channel flows under physical conditions representative of solar receivers.
		Solving the low-Mach number Navier-Stokes equations,
		T-LES results are evaluated a posteriori against Direct Numerical Simulation (DNS) data.
		We assess 12 subgrid-scale models.
		All models are based on the Anisotropic Minimum Dissipation (AMD) model.
		After computing a global error rate to evaluate all models,
		we select four for a detailed analysis regarding the effects of mesh resolution,
		numerical schemes, and model formulations.
		Results demonstrate that a two-layered mixed model combining the AMD/AMD-scalar with the Gradient model yields the best agreement with DNS.
	\end{abstract}

	\affiliation[promes]{organization={PROMES-CNRS, UPR 8521, Université de Perpignan Via Domitia},
		addressline={Rambla de la thermodynamique, Tecnosud},
		city={Perpignan},
		postcode={66100},
		country={France}}

	\affiliation[lisn]{organization={LISN--CNRS, UMR 9015, Université Paris-Saclay},
		addressline={Campus Universitaire d'Orsay, rue du Belvédère},
		city={Orsay},
		postcode={91405},
		country={France}}

\end{frontmatter}

\section{Introduction}
Next-generation solar power towers operate at higher temperatures than commercial molten salt plants, generating flows that are turbulent and highly anisothermal. The receiver is the key component of these plants. However, its large scale makes Direct Numerical Simulation (DNS) computationally prohibitive, as DNS must resolve every scale of turbulence. Alternatively, Thermal-Large Eddy Simulation (T-LES) resolves only the large scales while modeling the smallest scales through subgrid closures. T-LES allows for the evaluation of various receiver sections by modifying physical conditions. We model the flow using an asymmetrically heated channel, the simplest geometry possible.
This simplified geometry is representative of flows inside high temperature solar receiver. It allows to focus on the main physical phenomena: the coupling between turbulence and temperature. The various physical conditions that correspond to different distance from the solar receiver inlet are simulated thanks source terms in the Navier-Stokes equations~\citep{David_DNS_2023}.

Subgrid-scale tensor modeling has received attention in the computational fluid dynamics literature.
Since the creation of the field, substantial research has sought to define physical constraints for the subgrid-scale approach.
Galilean invariance~\citep{Speziale_1985,oberlack1997invariant,Pope_2000}, near-wall behavior~\citep{Chapman_Kuhn_1986,JIMENEZ1999252,Jimenez2013nearwall}, with models specifically tailored to behave similar near-wall scaling behavior~\citep{nicoud_subgridscale_1999,nicoud_using_2011,Trias2015buildingproper}, dissipation and realizability properties~\citep{Vreman_Geurts_Kuerten_1994}. For a more comprehensive overview of these physical constraints, see the review by~\citet{Silvius2017physicalconsistency}.

The most widely studied models assume that closure terms can be computed from resolved quantities.
These models can be classified as either functional or structural~\citep{sagautpierre_large_}.
Functional models, or eddy-viscosity models, assume the effects of the small scales to be purely dissipative,
analogous to viscous diffusion~\citep{boussinesq1877essai}.
While generally accurate and easy to implement, they can be overly dissipative.
In contrast, structural models mimic the mathematical structure of the closure term without imposing any physical assumption.
Because these models allow for backward energy transfer, they can be numerically unstable.

Numerous structural and functional models have been proposed in the literature.
A detailed overview is provided by~\citet{sagautpierre_large_}.
\rewriteFrancoise{Many functional models are based on the Smagorinsky model exist~Smagorinsky~\cite{Smagorinsky1963generalcircuit}, such as the dynamic Smagorinsky }{Many functional models are based on the work of~\cite{Smagorinsky1963generalcircuit}, such as the dynamic Smagorinsky} model~\citep{Germano1991dynamicsubgrid}, the Lilly dynamic constant Smagorinsky model~\cite{Lily1992aproposed} and numerous other variations of the Smagorinsky model~\citep{Iliescu2003largeeddy,ZHOU2019104319,rozema_minimumdissipation_2015,Trias2015buildingproper,ABBA2003521}.
Other functional models include the S3PQR model~\citet{Trias2015buildingproper}, the Anisotropic Minimum Dissipation (AMD) model by~\citet{rozema_minimumdissipation_2015}, and compressible and scalar variations of the AMD model~\citep{akbar2016amds,David2023TLES}.
Notable structural models include deconvolution models~\citet{HICKEL2006413,vonKaenel2002,Stolz1999a}, the Bardina model~\citep{bardina_improved_1980},
the scale similarity model and its variations ~\citep{Liu_Meneveau_Katz_1994,David2023TLES},
and the gradient model~\cite{LEONARD1975237}.
Some functional models incorporate structural modeling tools. For instance, the AMD models use the gradient model as a base, and the dynamic constant Smagorinsky model proposed by~\citet{ABBA2003521} uses the scale similarity model to determine the model constant.

The mixed models are a third type of model that combines the robustness of functional models with the structure and anisotropy of the structural models.
These mixed models were introduced by~\citet{bardina_improved_1980}.
In the literature, mixed models often use the~Smagorinsky~\cite{Smagorinsky1963generalcircuit} model, applying either fixed or dynamic coefficients to closure terms~\citep{Vreman1994ontheformulation,Vreman1996largeeddy,VREMAN_GEURTS_KUERTEN_1997,Zhand1993adynamic}.
More recently,~\citet{streher_mixed_2021} proposed a mathematical formulation for two-layered mixed models that accounts for near-wall flow phenomena, basing their closure on the AMD and Bardina models.
Two-layered mixed models introduce piecewise continuous functions instead of constants to scale the closures.
In the near-wall region,
the functional model constant is at a maximum, and its coefficient decreases towards the middle of the channel.
This formulation captures the near-wall domain characterized by dissipative energy exchanges~\cite{dupuyEnergyExchanges2018}.
The structural model constant does not change and has a naturally low amplitude close to the wall.
~\citet{streher_mixed_2021} obtained good results for different isothermal channel flows at various friction Reynolds numbers $Re_\tau=180, 395, 590$ and $950$.
\begin{rewriteAdrien}
	While all the aforementioned works concentrate strictly on isothermal flows, \citet{David2023TLES} evaluated two-layered mixed models with asymmetrical heating. In their work, they utilized the AMD and Bardina model combination for both the momentum and energy closures. They showed that an increase in imposed heat flux degrades model performance, illustrating that these specific closures do not accurately capture the coupling between the turbulence and thermal components of the flow, and are prone to numerical instabilities.

	To address these limitations, this work expands upon the configurations tested by \citet{David2023TLES} by introducing and evaluating twelve new T-LES models. Specifically, this study differentiates itself by:
	\begin{enumerate}
		\item Replacing the standard AMD closure in the energy equation with the AMD-scalar model to better capture thermal dissipation.
		\item Changing the Bardina structural model with the Gradient model within the two-layered mixed to improve velocity-temperature coupling.
		\item Quantifying the influence of different numerical schemes (2nd-order, 4th-order, and QUICK) on the accuracy and stability of these new scalar and compressible AMD formulations.
	\end{enumerate}
\end{rewriteAdrien}

This paper is organized as follows.
Section~\ref{label-pa} details the governing equations, geometry, and numerical setup.
The different types of subgrid-scale models are presented in section~\ref{sec_subgrid_models}. The results are discussed and analyzed in section~\ref{sec_results}, and conclusions are drawn in section~\ref{sec_conclusion}.
\section{Numerical setting}
\label{label-pa}
This section describes the low-Mach number Navier-Stokes equations proposed by~\citet{paolucci_filtering_1982}.
This formulation gives provides a middle ground between the incompressible and compressible Navier-Stokes equations.
It accounts for large temperature-driven density variations while neglecting acoustic wave propagation.
This enables the use of numerical tools designed for incompressible flows.
The pressure is divided into the thermodynamic pressure, constant in space $P_0(t)$, and the mechanical pressure.
We apply the Stokes hypothesis~\citet{Papalexandris2019}.
As suggested by~\citet{dupuy_2019_isothermal}, we employ the Favre formulation to filter the Navier-Stokes equations.
For any quantity $\phi$, its filtered version is $\fa{\phi}=\f{\rho\phi}/\f{\rho}$, where $\f{\cdot}$ denotes the classical volumetric filter.
Although multiple nonlinear terms arise from the filtering process, we retain only
the subgrid terms responsible for the nonlinearity of the momentum convection
and the mass-velocity correlation, as suggested by~\citet{Dupuy2019_a_priori}.
\begin{itemize}
	\item Mass conservation equation
	      \begin{equation}
		      \der{\f{\rho}}{t} + \der{\f{\rho} \fa{U_{j}}}{x_j} = 0,
	      \end{equation}
	\item Momentum conservation equation
	      \begin{equation}
		      \begin{aligned}
			      \der{\f{\rho} \fa{U}_{\!i}}{t} = - \der{\left(\smash[t]{\f{\rho} \fa{U}_{\!j} \fa{U}_{\!i} + \f{\rho} G_{U_j U_i}}\right)}{x_j} - \der{\f{P}}{x_i} + \der{\varSigma_{ij}({\fa{U}},\fa{T})}{x_j},
		      \end{aligned}
	      \end{equation}
	\item Energy conservation equation
	      \begin{equation}
		      \der{}{x_j}\left(\fa{U}_{\!j} + \f{\rho} G_{U_j/\rho}\right) = - \frac{1}{\gamma P_{0}}\left( (\gamma - 1)\left[\der{Q_j(\fa{T})}{x_j} - H_s\right] + \der{P_{0}}{t} \right),
		      \label{eq_energy_conserv}
	      \end{equation}
	\item Ideal gas law
	      \begin{equation}
		      \fa{T} = \frac{P_{0}}{\f{\rho} r},
		      \label{eq_perf_gas}
	      \end{equation}
\end{itemize}
where $\rho$ is the density, $T$ the temperature, $\gamma$ the heat capacity ratio,
$r$ the gas specific constant, $t$ the time, $P$ the mechanical pressure,
$P_0$ the thermodynamical pressure, $U_i$ the velocity in the $i$-th direction,
$x_i$ the coordinate in $i$-th direction and $H_s$ the heat sink source term. We use the Einstein summation convention.
The functions $\varSigma_{ij}(\vv{U}, T)$ and $Q_j(T)$ are used to compute the
shear-stress tensor and conductive heat flux associated with a given velocity
and temperature.
We assume a Newtonian fluid and Fourier's law,
\begin{align}
	\varSigma_{ij}(\vv{U}, T) ={} & \mu(T) \left(\der{U_i}{x_j} + \der{U_j}{x_i}\right) - \frac{2}{3} \mu(T) \der{U_k}{x_k} \delta_{ij}, \\
	Q_j(T) {}                     & = - \lambda(T) \der{T}{x_j},
\end{align}
with $\mu$ the dynamic viscosity, $\lambda$ the thermal conductivity and
$\delta_{ij}$ the Kronecker delta.

We define the momentum convection closure term as
$\smash[t]{G_{U_j U_i} ={} \fa{U_j U_i} - \fa{U}_{\!j} \fa{U}_{\!i}}$.
The density-velocity correlation closure term is defined as
$\smash[t]{G_{U_j/\rho} ={} \fa{U_j/\rho} - \fa{U}_{\!j}/\f{\rho}}$.
Using equation~\ref{eq_perf_gas}, the equivalence can be drawn $\f{\rho} \smash[t]{G_{U_j/\rho}=\left[\fa{U_j T} -\fa{U_j}\fa{T}\right]/\fa{T}=G_{U_j T}/\fa{T}}$.

\begin{rewriteAdrien}
	The heat-carrying fluid is pressurized air. Its dynamic viscosity $\mu(T)$ is computed using Sutherland's law~\citep{sutherland1893lii}
	\begin{equation}
		\mu(T) = \mu_0 \left(\frac{T}{T_0}\right)^{\frac{3}{2}} \frac{T_0 + S}{T + S},
	\end{equation}
	with $\mu_0 = \SI{1.716e-5}{\pascal\second}$, $S=\SI{110.4}{\K}$ and the reference temperature $T_0 = \SI{273.15}{\K}$.
	The thermal conductivity $\lambda(T)$ is computed using a similar Sutherland formulation
	\begin{equation}
		\lambda(T) = \lambda_0 \left(\frac{T}{T_0}\right)^{\frac{3}{2}} \frac{T_0 + S}{T + S},
	\end{equation}
	where the reference thermal conductivity is ${\lambda_0 = \SI{0.0261}{\watt\per\meter\per\K}}$.
	The specific gas constant for air is ${r=\SI{287}{\joule\per\kilogram\per\K}}$, and the heat capacity ratio is assumed constant at $\gamma = 1.4$.
\end{rewriteAdrien}
\subsection{Geometry and mesh}
To model flows inside next-generation solar receivers, we use a bi-periodic channel, which represents the simplest possible geometry. Because the streamwise ($x$) and spanwise ($z$) directions are periodic, the grid spacing along these axes is uniform.
In the wall-normal direction, the mesh is finer at the wall and coarser at the center of the channel.
The distance from the wall ($y$) follows a hyperbolic tangent law:
\begin{equation}
	y_k = L_y \Big( 1 + \frac{1}{a} \tanh \Big[ \Big( \frac{k-1}{N_y -1} \Big) \tanh ^{-1} (a) \Big]
	\Big )
	\label{eq_mesh_wall_norm}
\end{equation}
where $a$ is the mesh dilatation coefficient, and $N_y$ is the number of wall-normal grid points. The computational domain is illustrated in~\cref{fig_geometry}.
\begin{NoHyper}
	\begin{figure*}
		\begin{center}
			\includegraphics[width=0.85\textwidth]{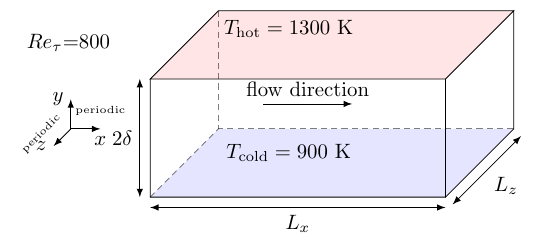}
		\end{center}
		\caption{Representation of the simulation domain}
		\label{fig_geometry}
	\end{figure*}
\end{NoHyper}
The present simulations are performed using TRUST/TrioCFD, developed by the French Alternative Energies and Atomic Energy Commission~\citep{calvin_objectoriented_2002}.
It has been validated across numerous fluid flow simulations~\citep{dupuy_posteriori_2019a, Toutant2018, Avellaneda2019, Dupuy2019, aulery_spectral_2017, david_numerical_2021, David2021, david_investigation_2021, David2023TLES}.
The code employs a finite difference method on a staggered grid and advances in time using a third-order Runge-Kutta scheme~\citet{Runge1895, Kutta1901}.
For spatial discretization, momentum convection is evaluated using a second- or fourth-order centered scheme, while mass convection utilizes either a second-order centered scheme or a third-order QUICK scheme. Finally, velocity divergence and temperature diffusion are computed using a second-order centered scheme.
\begin{table*}[htbp]
	\begin{center}
		\begin{tabular}{*{3}{c}}
			\toprule 
			Mesh & $N_x\times N_y \times N_z$  & $\Delta_x^+; \Delta_y^+; \Delta_z^+$ \\
			\midrule 
			DNS  & $1152\times 746 \times 768$ & $10.6; 0.4$---$5.3 ; 5.3$            \\
			\midrule 
			A    & $256 \times 152 \times 192$ & $47.5; 1.0$---$31.2; 21.1$           \\
			\midrule 
			B    & $192 \times 152 \times 128$ & $63.3; 1.0$---$31.2; 31.7$           \\
			\midrule 
			C    & $160 \times 152 \times 96 $ & $76.0; 1.0$---$31.2; 42.2$           \\
			\bottomrule 
		\end{tabular}
	\end{center}
	\caption{Simulation parameters.}
	\label{tab_params}
\end{table*}
\subsection{Subgrid-scale models}
\label{sec_subgrid_models}
The velocity-velocity and velocity-mass closures are approximated as
\begin{align}
	\label{eq_closure}
	G_{U_j U_i}    & \approx \tau_{ij}^{\mathrm{mod}}(\vv{\fa{U}}, \f{\Delta}),       \\
	G_{U_j / \rho} & \approx \pi_{j}^{\mathrm{mod}}(\vv{\fa{U}}, 1/\rho, \f{\Delta}),
\end{align}
where $\tau_{ij}^{\mathrm{mod}}$ and $\pi_j^{\mathrm{mod}}$ represent the subgrid-scale models selected to model highly anisothermal flows. The filter width is defined as $\f{\Delta}=\sqrt[3]{\f{\Delta}_x\f{\Delta}_y\f{\Delta}_z}$. This study exclusively evaluates fixed-constant models, as the performance of dynamic-constant models does not exceed that of their fixed-constant counterparts. Furthermore, dynamic models incur higher computational costs~\citep{dupuy_posteriori_2019a,dupuy_2019_isothermal}.
Following the methodology of~\citet{David2023TLES}, we explore the effects of varying these  constants within a mixed model context.

To evaluate the subgrid closures under different thermal loads, two distinct operating conditions are defined by varying the heat sink term $H_s$ in equation~\ref{eq_energy_conserv}. The baseline condition features no heat sink ($H_s = 0$), which corresponds to a hot-side boundary heat flux of $\SI{98}{\kilo\watt\per\meter\squared}$.. Conversely, condition S1 introduces a uniform volumetric heat sink of $\SI{55}{\mega\watt\per\meter\cubed}$. This added source term alters the temperature profiles and drives a 2.5-fold increase in the hot-side boundary heat flux, raising it to $\SI{263}{\kilo\watt\per\meter\cubed}$.
\subsubsection{Functional models}
Functional models encompass family of eddy-viscosity models based on the Boussinesq hypothesis~\cite{boussinesq1877essai}, which assumes that the effects of small scales is purely dissipative. This assumption often yields overly dissipative models, as energy is strictly removed from the resolved scales. The closure tensor for the velocity-velocity correlation is expressed as
\begin{equation}
	\tau_{ij}^{\mathrm{mod}}(\vv{\fa{U}}, \f{\Delta})  = -2 \nu_e^{\mathrm{mod}}(g,d,\f{\Delta})S_{ij},
	\label{eq_sgs_tau}
\end{equation}
where $S_{ij}=\frac{1}{2}\left(\pfrac{\fa{U_i}}{x_j}+\pfrac{\fa{U_j}}{x_i}\right)$ is the deformation rate tensor, $g$ is the velocity gradient $g_{ij}=\pfrac{\fa{U_i}}{x_j}$, and $\nu_e^{\mathrm{mod}}$ is the turbulent eddy-viscosity specific to the chosen model. This turbulent eddy-viscosity can also approximate the velocity-mass term by introducing of the turbulent Prandtl number $Pr_t$
\begin{equation}
	\pi_j^{\mathrm{mod}}(\vv{\fa{U}},s,\f{\Delta})=-\frac{\nu_e^{\mathrm{mod}}(g,d,\f{\Delta})}{Pr_t}d_j,
	\label{eq_pi}
\end{equation}
where $d_j=\pfrac{s}{x_j}$ is the scalar gradient for any scalar $s$. Following the recommendation of~\citet{David_DNS_2023}, the turbulent Prandtl number is fixed at $Pr_t=0.9$.

We employ the Anisotropic Minimum Dissipation (AMD) model proposed by~\citet{rozema_minimumdissipation_2015}, defined as
\begin{equation}
	\nu_e^{\mathrm{AMD}}(g,d,\f{\Delta})=C^{\mathrm{AMD}}\frac{\max(0, -G_{ij} S_{ij})}{g_{kl}^2}
	\label{eq_amd_mod}
\end{equation}
where $G_{ij} = \f{\Delta_k}^2 g_{ik}g_{jk}$ is the base gradient model.

We also evaluate two other eddy-viscosity models based on the AMD model. The first one is the scalar AMD model ($\mathrm{AMD}^\mathrm{s}$), proposed by~\citet{akbar2016amds}
\begin{equation}
	\nu_e^{\mathrm{AMD}^\mathrm{s}}(g, d, \f{\Delta}) = C^{\mathrm{AMD}} \frac{\max{(0, -D_j d_j)}}{d_m^2},
	\label{eq_amd_scalar}
\end{equation}
where $D_j=\f{\Delta_k}^2 g_{jk} d_k$ is the scalar gradient model. The second is the compressible AMD model ($\mathrm{AMD}^\mathrm{c}$) proposed by~\citet{David2023TLES}, which accounts for compressibility through the trace of the deformation rate tensor $S_{ij}$
\begin{equation}
	\nu_e^{\mathrm{AMD}^{\mathrm{c}}}(g,d,\f{\Delta})=C^{\mathrm{AMD}^\mathrm{c}}\frac{\max(0,-(G_{ij}-\frac{1}{3}G_{kk}\delta_{ij})S_{ij})}{(S_{lm} - \frac{1}{3}S_{kk}\delta_{km})S_{lm}}.
	\label{eq_amd_c}
\end{equation}
\subsubsection{Structural models}
Structural models assume that
the closure term can be approximated using the same mathematical structure as the real closures,
without imposing specific physical assumptions about the flow.
Unlike purely dissipative functional models, this approach allows for backward energy transfer, although it renders structural models prone to numerical instabilities.
The Bardina model (Bard) proposed by~\citet{bardina_improved_1980} is expressed as
\begin{align}
	\tau_{ij}^{\mathrm{Bard}} & =C^{\mathrm{Bard}}\Big( \fa{U_j}\fa{U_i} - \ff{\fa{U_j}}\ff{\fa{U_i}} \Big), \\
	\pi_j^{\mathrm{Bard}}     & =C^{\mathrm{Bard}} (\fa{U_j}\fa{T} - \ff{\fa{U_j}}\ff{\fa{T}}),
	\label{eq_bard}
\end{align}
\begin{rewriteAdrien}
	By substituting the discrete one-cell (two-node) primary filter and the three-cell (four-node) combined test filter into the Bardina closures, the explicit equations are obtained.

	For the diagonal terms of the subgrid tensor, filtered in the streamwise direction ($n$):
	\begin{align}
		\tau_{ii}^{\mathrm{Bard}} & = C^{\mathrm{Bard}} \left[ \frac{1}{4} (U_i^n + U_i^{n+1})^2 \right. \nonumber    \\
		                          & \quad \left. - \frac{1}{16} (U_i^{n-1} + U_i^n + U_i^{n+1} + U_i^{n+2})^2 \right]
	\end{align}

	For the extra-diagonal terms, preserving the cross-directional filtering where $n$ and $m$ are the streamwise and spanwise node indices, respectively:
	\begin{align}
		\tau_{ij}^{\mathrm{Bard}} & = C^{\mathrm{Bard}} \Bigg[ \frac{1}{4} (U_i^{m-1} + U_i^m)(U_j^{n-1} + U_j^n) \nonumber \\
		                          & \quad - \frac{1}{16} (U_i^{m-2} + U_i^{m-1} + U_i^m + U_i^{m+1}) \nonumber              \\
		                          & (U_j^{n-2} + U_j^{n-1} + U_j^n + U_j^{n+1}) \Bigg]
	\end{align}

	For the subgrid temperature flux in the spanwise direction ($j$), the fully symmetric formulation applies the filters across the corresponding spanwise nodes ($m$):
	\begin{align}
		\pi_j^{\mathrm{Bard}} & = C^{\mathrm{Bard}} \Bigg[ \frac{1}{2} U_j(T^{m-1} + T^m) \nonumber \\
		                      & \quad - \frac{1}{4} U_j(T^{m-2} + T^{m-1} + T^m + T^{m+1}) \Bigg]
	\end{align}
\end{rewriteAdrien}
The filtering procedure for the Bardina model follows the methodology proposed by~\citet{streher_mixed_2021}.

The scale similarity model (Sim) proposed by~\citet{Liu_Meneveau_Katz_1994} is defined as
\begin{align}
	\label{eq_ssim}
	\tau_{ij}^{\mathrm{sim}} & =C^{\mathrm{sim}}\Big(\ff{\fa{U_j}\fa{U_i}} - \ff{\fa{U_j}}\ff{\fa{U_i}}\Big), \\
	\pi_{j}^{\mathrm{sim}}   & =C^{\mathrm{sim}}\Big(\ff{\fa{U_j}\fa{T}} - \ff{\fa{U_j}}\ff{\fa{T}}\Big),
\end{align}
where $\ff{\cdot}$ denotes the top-hat filter.
Its compressible variant, proposed by~\citet{David2023TLES}, takes the form
\begin{align}
	\label{eq_ssim_c}
	\tau_{ij}^{\mathrm{sim}^c} & =C^{\mathrm{sim}^c}\Big(\ff{\f{\rho}\fa{U_j}\fa{U_i}} - \ff{\f{\rho}\fa{U_j}}\ff{\f{\rho}\fa{U_i}}/\f{\rho}\Big), \\
	\pi_{j}^{\mathrm{sim}^c}   & =C^{\mathrm{sim}^c}\Big(\ff{\f{\rho}\fa{U_j}\fa{T}} - \ff{\f{\rho}\fa{U_j}}\ff{\f{\rho}\fa{T}}/\f{\rho}\Big).
\end{align}

Finally, the gradient model (Grad) proposed by~\citet{LEONARD1975237} is written as
\begin{align}
	\tau_{ij}^{\mathrm{Grad}} = \frac{C^{\mathrm{Grad}}}{12}\f{\Delta_k}^2 g_{ik}g_{jk}, \\
	\pi_{j}^{\mathrm{Grad}}   = \frac{C^{\mathrm{Grad}}}{12}\f{\Delta_k}^2 g_{jk}d_k.
\end{align}
The gradient model is based on a Taylor series expansion of the filter. For an even filter kernel, a filtered quantity $\f{\phi}=\phi\ast G$ is approximated as:
\begin{equation}
	\f{\phi}=\phi\ast G = \sum_{n=1}^{\infty} \gamma_{2n} \frac{\partial^{2n} \phi}{\partial x_k^{2n}},
\end{equation}
with $G$ is the spatial convolution filter, and where the coefficients $\gamma_n$ depend on the filter type (e.g., top-hat or Gaussian). Applying the Van Cittert iterative deconvolution provides a series expansion of the inverse filter:
\begin{equation}
	G^{-1} = \sum_{n=1}^{\infty} (1 - G)^n.
\end{equation}
Using these expansions to approximate the subgrid-scale tensor and truncating at the second order yields the base formulation. We generalize this definition by introducing the constant $C^{\mathrm{Grad}}=1$. A more detailed mathematical derivation is shown in~\citet{dupuy_analyse_2018} section 7.2.1.
\subsubsection{Mixed models}
Mixed models combine the features of functional and structural models. Because functional models tend to overly dissipate energy and structural models are prone to numerical instability, a balance can be struck between the two.
To this end, we evaluate several mixed models and introduce the concept of model layering. A one-layered mixed model is defined as
\begin{align}
	\label{eq_mixed_model_one_l}
	\tau_{ij} & =\alpha_1 \tau_{ij}^\textrm{func}+\beta_1\tau_{ij}^\textrm{struct}, \\
	\pi_{j}   & =\alpha_2 \pi_{j}^\textrm{func}+\beta_2\pi_{j}^\textrm{struct},     %
	\label{eq:constant_models}
\end{align}
where $\alpha_1,\alpha_2,\beta_1,\beta_2$ are constants.

\begin{figure*}
	\begin{center}
		\includegraphics[width=0.5\textwidth]{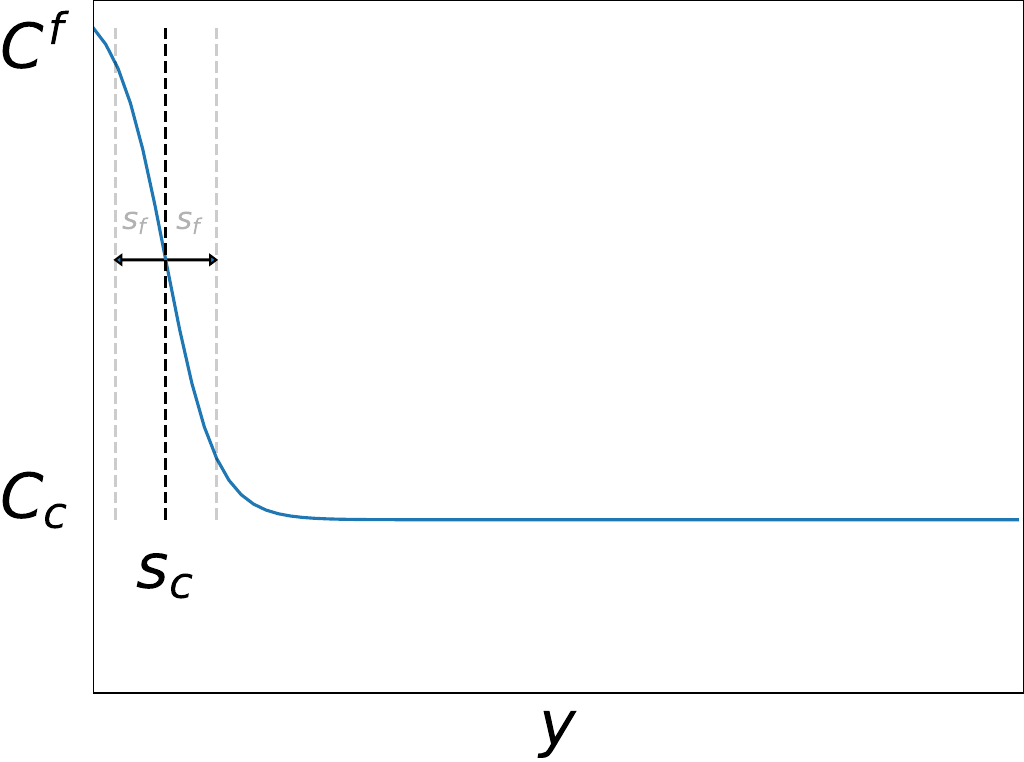}
	\end{center}
	\caption{Functional constant evolution as a function of the height in the channel.}
	\label{fig_constant_func}
\end{figure*}

Two-layered mixed models vary the functional model coefficient as a function of wall distance, while the structural model coefficient remains constant. This is translated by $\alpha_1=\alpha_1(y)$, $\alpha_2=\alpha_2(y)$.
In the viscous sublayer,
viscous dissipation dominates energy exchanges,
therefore, the functional model contribution is maximized.
Closer to the center of the channel, energy exchanges are dominated by turbulent structures,
requiring a reduced functional model coefficient.
In this outer region, the structural model adequately approximates the closure.
The functional model constant is defined as
\begin{equation}
	C^\mathrm{func, \ dyn}(y) = C^\mathrm{func} + \left( \frac{1}{2} + \frac{1}{2} \tanh{\Big( \frac{y-s_c}{s_f}\Big)}\right)(C^\mathrm{center} - C^\mathrm{func}),
\end{equation}
where $y$ is the wall-normal height, $C^\mathrm{center}$ is the coefficient at the channel center, and $C^\mathrm{func}$ is the value at the boundary. A visualisation of the constant variation from the wall to the middle of the channel is given in figure ~\ref{fig_constant_func}.
Following the work of~\citet{streher_mixed_2021}, and the methodology outlined by~\citet{David2023TLES}, the smoothing center and smoothing factor are set to $s_f=0.00016252$, and $s_c=0.00023217$ respectively.
These constants target second-order velocity fluctuation peak in the near wall region to keep the functional model at a maximum close to the wall, and at a minimum in the outer region.
This transition corresponds to the interface height $y_{\mathrm{int}}$, which is 4\% of the total height of the canal, and ${s_f=0.7\cdot s_c}$ according to~\citet{streher_mixed_2021}. 
\section{Results}
\label{sec_results}
\subsection{Data gathering and averaging}
Data obtained from simulations are averaged after full statistical convergence.
The averaging operation performed over the streamwise direction $x$, the spanwise direction $z$, and time $t$
\begin{equation}
	\label{eq_average_operator}
	\av{\phi}(y) = \frac{1}{L_x L_z (t_b-t_a)}\intl_{0}^{L_x}\intl_{0}^{L_z}\intl_{t_a}^{t_b}\phi(x,y,z,t) \diff x \diff z \diff t.
\end{equation}
Second-order terms, such as the Reynolds stress tensor require the addition of their closure terms for comparison with DNS data.
Assuming~$\av{\phi}\approx\av{\fa{\phi}}$, this decomposition is derived as
\begin{align}
	R_{ij}^\mathrm{DNS} & =\av{U_jU_i} - \av{U_j} \av{U_i},                       \\
	R_{ij}^\mathrm{DNS} & \approx \av{\fa{U_jU_i}} - \av{\fa{U_j}}\av{\fa{U_i}},  \\
	R_{ij}^\mathrm{DNS} & = \av{\fa{U_jU_i}} - \av{\fa{U_j}}\av{\fa{U_i}}
	+\av{\fa{U_i}\fa{U_j}}
	-\av{\fa{U_i}\fa{U_j}},                                                       \\
	R_{ij}^\mathrm{DNS} & =  \underbrace{\av{\fa{U_i}\fa{U_j}}
		- \av{\fa{U_j}}\av{\fa{U_i}}}_{R_{ij}^\mathrm{LES}}
	+ \underbrace{\av{\fa{U_j U_i}}
	-\av{\fa{U_i}\fa{U_j}}}_{\av{\tau^{\mathrm{mod}}_{ij}}},                      \\
	R_{ij}^\mathrm{DNS} & =  R_{ij}^\mathrm{LES} + \av{\tau^{\mathrm{mod}}_{ij}},
	\label{eq_proof_rms}
\end{align}
The diagonal components of the Reynolds tensor are evaluated using their deviatoric parts
\begin{align}
	R_{ii}^{\mathrm{DNS,dev}} & = \av{U_i^2}-\av{U_i}^2 - \sum_{k=1}^{3}\frac{1}{3}(\av{U_{k}^2}-\av{U_k}^2),                  \\
	                          & = R_{ii}^\mathrm{LES,dev} + \av{\tau_{ii}^\mathrm{mod}(\vv{\fa{U}}, \f{\Delta})}^\mathrm{dev}.
	\label{eq_rms_dev}
\end{align}
The deviatoric part of the Reynolds stress tensor for any diagonal term is the correlation term without the third of the trace.
Similarly, this comparison is also applied to temperature-velocity correlations
\begin{equation}
	\label{eq_comp_temp_vel}
	\underbrace{\av{U_j T} - \av{U_j} \av{T} }_{R_{jT}^\mathrm{DNS}}=\underbrace{\av{\fa{U_j}\fa{T}} - \av{\fa{U_j}}\av{\fa{T}}}_{R_{jT}^\mathrm{LES}} + \underbrace{\av{\pi_j^\mathrm{mod}}}_{\text{model}}.
\end{equation}
We use the local scaling $\cdot^+$ defined as
\begin{align}
	x_i^+          & =x_i Re_\tau/h,                      \\
	U_i^+          & =\frac{U_i}{U_\tau},                 \\
	\av{R_{ij}}^+  & =\frac{\av{R_{ij}}}{U_\tau^2},       \\
	\av{U'_i T'}^+ & =\frac{\av{U'_i T'}}{U_\tau T_\tau},
	\label{eq_adim}
\end{align}
where $Re_\tau=\frac{U_\tau h}{\nu}$ is the friction Reynolds number, ${U_\tau=\sqrt{\nu_w\pfrac{U_x}{ y}|_{w}}}$ is the friction velocity, and $T_\tau=\frac{\av{\lambda\pfrac{T}{y}}_{|w}}{\rho C_p U_\tau}$ is the friction temperature, and $C_p$ is the specific heat capacity at constant pressure.
Additionally, the Nusselt number is computed using
\begin{equation}
	Nu=\frac{D_h \langle \lambda\pfrac{T}{y} \rangle_{|w}}{\lambda_w (T_w - T_b)},
	\label{eq_nu}
\end{equation}
where $D_h=4h$ is the hydraulic diameter for a bi-periodic channel, $T_w$ the boundary temperature, and $T_b=\frac{\intl_{0}^{2h}\langle \rho U_x T \rangle (y) \diff y}{\intl_{0}^{2h}\langle \rho U_x \rangle (y) \diff y}$ is the bulk temperature.
The friction coefficient $C_f$ is given by
\begin{equation}
	C_f=\frac{2\mu_w \av{\frac{\partial U_x}{\partial y}}_{|w}}{\rho_b U_b^2},
	\label{eq_cf}
\end{equation}
where $\rho_b=\frac{1}{2h}\intl_{0}^{2h}\langle \rho \rangle(y) \diff y$ is the bulk density and $U_b=\frac{\intl_{0}^{2h}\langle U_x \rho \rangle(y) \diff y}{\intl_{0}^{2h}\langle \rho \rangle(y) \diff y}$ the bulk velocity.
\subsection{Error quantification using DNS statistics}
\rewriteFrancoise{To quantify the accuracy of the subgrid-scale models, we compare selected profiles from the LES and DNS.}{To quantify the accuracy of the subgrid-scale models, we compare the wall-normal distributions of statistical quantities predicted by the LES against the reference DNS.} First, the DNS profile is interpolated onto the LES mesh. Then, a point-to-point relative error is computed for a given channel half as
\begin{equation}
	\varepsilon^{\mathrm{LES}, \ i}_\phi=\frac{\suml_{n=1}^{N_y/2} \log{\Big(\frac{y_{n+1}}{y_n}\Big)} \Big| (\av{\phi}^\mathrm{DNS}(y_n) - \av{\phi}^\mathrm{LES}(y_n)) \av{\phi}^\mathrm{DNS}(y_n) \Big|}{\suml_{n=1}^{N_y/2} \log{\Big(\frac{y_{n+1}}{y_n}\Big)} \av{\phi}^{\mathrm{DNS}^2} },
	\label{eq_err_half}
\end{equation}
where $\phi$ represents any of the considered first-order statistics ($U$, $V$ and $T$), or second-order statistics ($U'^2$, $V'^2$, $W'^2$, $U'V'$, $U'T'$ and $V'T$). The error on the Nusselt number $Nu$ and the friction coefficient $C_f$ are computed using the scalar relative errors
\begin{align}
	\varepsilon^\mathrm{LES}_{Nu}  & =\frac{|Nu^\mathrm{LES} - Nu^\mathrm{DNS}|}{Nu^\mathrm{DNS}},    \\
	\varepsilon^\mathrm{LES}_{C_f} & =\frac{|C_f^\mathrm{LES} - C_f^\mathrm{DNS}|}{C_f^\mathrm{DNS}}.
\end{align}
This error function is similar to the one proposed by~\citet{David2023TLES}. The mean and second-order errors are then normalized by the maximum error observed across all models
\begin{equation}
	\varepsilon^{\mathrm{LES}, \ i}_\mathrm{mean}=\frac{\suml_\phi \varepsilon_\phi^{\mathrm{LES}, \ i}}{\max{\Big(\suml_\phi \varepsilon_\phi^\mathrm{LES}\Big)}}, \quad
	\varepsilon^{\mathrm{LES}, \ i}_\mathrm{rms}=\frac{\suml_\phi \varepsilon_\phi^{\mathrm{LES}, \ i}}{\max{\Big(\suml_\phi \varepsilon_\phi^\mathrm{LES}\Big)}}.
	\label{eq_err_sum}
\end{equation}
The subscript \texttt{rms} corresponds to the errors associated with second-order turbulent correlations ($U_i'U_j'$, $U_i' T'$). The total error is then computed as a weighted sum, scaled by the number of variable in each category:
\begin{equation}
	\varepsilon_{\mathrm{tot}}^{\mathrm{LES}, \ i}=\frac{n\varepsilon_{\mathrm{mean}}^{\mathrm{LES}, \ i}+m\varepsilon_{\mathrm{rms}}^{\mathrm{LES}, \ i}}{n+m},
	\label{eq_err_tot}
\end{equation}
where $n$ is the number of mean quantities, and $m$ is the number of correlations.
The error on the second-order temperature correlation $\av{T'^2}$ is not taken into account as no closure term exists.
\subsection{Error quantification of T-LES}
Table~\ref{tab_models} summarizes the configurations evaluated in this study, which exclusively considers two-layered mixed models and functional models.
All mixed models have a channel center constant of $C_c=0.15$.
Five mixed models and seven functional models are presented.
Among the mixed models, two use the AMD scalar closure instead of the AMD model for the $G_{U/\rho}$ closure, while the remaining cases use the Gradient model.
Finally, the functional models assess the performance of the AMD compressible---AMD scalar, and the AMD compressible---AMD compressible combination with different numerical schemes, including the QUICK scheme.

\begin{figure*}[htbp]
	\begin{center}
		\includegraphics[width=\textwidth]{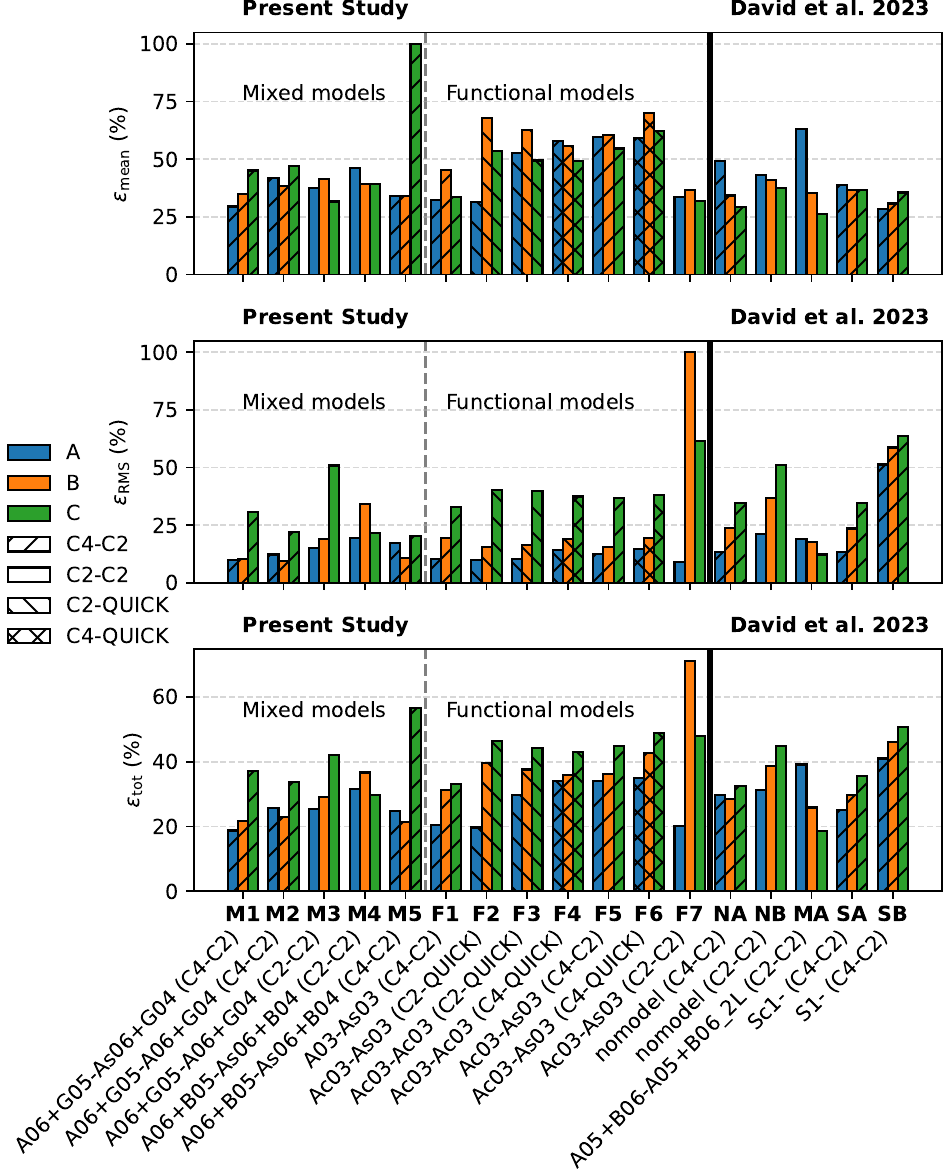}
	\end{center}
	\caption{Normalized errors for the 12 tested models.
		The top panel represents the mean error, the middle one, the RMS error, and the last one the weighted average of both errors.
		Meshes A, B and C represent the finest to coarsest meshes.
		The remaining 6 models on the right are the best performing models taken from~\cite{David2023TLES} as a point of comparison.
		The short hand names are also written above each model.}
	\label{fig_error_bar}
\end{figure*}

Figure~\ref{fig_error_bar} displays the error rates computed using equation~\ref{eq_err_sum}, and~\ref{eq_err_tot}.
For a more exhaustive representation of the error rates, the reader is referred to the appendix~\ref{sec:annexe_error_sec}.
Reference models from~\citet{David2023TLES} are positioned to the right of the F7 model. The top, middle, and bottom panels respectively display the mean, second-order and total error rates.
Models are sorted in ascending order of total error averaged over the simulated meshes.
We designate the overall performance by the total performance averaged over the simulated meshes, shown in table~\ref{tab:err_total}.
The best overall performing model is the M1 two-layered mixed model with 25.9\% error, with the M2 model in second with 27.4\%, almost matching the previous best model MA at 27.9\% error by~\citet{David2023TLES}.
The presented model M1 performs best on the A mesh and second best on the B mesh.
Regarding second-order statistics, the M2 model performs best (14.8\% error) with models M1 (17\%), M5 (16.1\%), and MA (16.4\%) performing similarly with error percentage differences of 2.2\% at most.
The mean statistics are best captured by the SB structural model with an error rate of 31.7\%, followed by the F7 (34\%) and M1 (36.5\%) models.
The M1 model utilizes a C4---C2 numerical scheme with constants at the wall $\tau_{ij}=0.6\tau_{ij}^{\mathrm{AMD}} + 0.5\tau_{ij}^{\mathrm{Grad}}$ and $\pi_j=0.6\pi_j^{\mathrm{AMD^s}} + 0.4\pi_j^\mathrm{Grad}$.
The MA model utilizes a C2---C2 numerical scheme with constants at the wall $\tau_{ij}=0.6\tau_{ij}^{\mathrm{AMD}} + 0.5\tau_{ij}^{\mathrm{Bard}}$ and $\pi_j=0.6\pi_j^{\mathrm{AMD}} + 0.4\pi_j^\mathrm{Bard}$.

The analysis of functional models highlights a coupling between model and numerical scheme, with no clear isolation of their individual effects.
While coarser meshes generally reduce accuracy for both mean and second-order statistics, this trend can be hidden by error compensation, where numerical diffusivity substitutes for the intended model diffusivity.
Furthermore, any configuration using the QUICK scheme exhibits poorer performances, which acts as an implicit filter.
Conversely, for functional models, F1 shows the best accuracy for both mean and total error ranking fourth overall, behind the M1, M2 and MA models.
Taking into account the compressibility in the momentum convection term degrades accuracy as demonstrated by the rest of the functional models (F2 to F7).

To analyze the wall-normal profiles,
the authors chose the MA, NA, M1, F1 models, to show differences in the studied mixed model and the best performing models from~\citet{David2023TLES}, the NA model for the best performing no-model T-LES, and the F1 model for the best performing functional model.
For the sake of clarity, only the finest and coarsest meshes (A) and (C) are shown for each simulation. Furthermore, to dissociate the hot from the cold sides, the hot side values have their sign inverted to to not add unnecessary clutter.
\begin{table*}
	\centering
	\resizebox{\textwidth}{!}{\begin{tabular}{*{11}{c}}
			\toprule
			Model name                  & shorthand & \multicolumn{2}{c}{$\tau$ model} & \multicolumn{2}{c}{$\pi$ model} & Functional                & \multicolumn{2}{c}{Numerical scheme}                                         \\
			                            & name      & $C \cdot$ Func.                  & $C \cdot$ Struct.               & $C \cdot$ Func.           & $C \cdot$ Struct.                    & Constant    & Mom. conv. & mass conv. \\
			\midrule
			\multicolumn{9}{l}{\textit{Present Study}}                                                                                                                                                                              \\
			A06+G05-As06+G04\_c4\_c2    & M1        & 0.6 AMD                          & 0.5 Grad                        & 0.6 $\text{AMD}^\text{s}$ & 0.4 Grad                             & 0.15        & c4         & c2         \\
			A06+G05-A06+G04\_c4\_c2     & M2        & 0.6 AMD                          & 0.5 Grad                        & 0.6 AMD                   & 0.4 Grad                             & 0.15        & c4         & c2         \\
			A06+G05-A06+G04\_c2\_c2     & M3        & 0.6 AMD                          & 0.5 Grad                        & 0.6 AMD                   & 0.4 Grad                             & 0.15        & c2         & c2         \\
			A06+B05-As06+B04\_c2\_c2    & M4        & 0.6 AMD                          & 0.5 Bard                        & 0.6 $\text{AMD}^\text{s}$ & 0.4 Bard                             & 0.15        & c2         & c2         \\
			A06+B05-As06+B04\_c4\_c2    & M5        & 0.6 AMD                          & 0.5 Bard                        & 0.6 $\text{AMD}^\text{s}$ & 0.4 Bard                             & 0.15        & c4         & c2         \\
			A03-As03\_c4\_c2            & F1        & 0.3 AMD                          & $\emptyset$                     & 0.3 $\text{AMD}^\text{s}$ & $\emptyset$                          & 0.3         & c4         & c2         \\
			Ac03-As03\_c2\_quick        & F2        & 0.3 $\text{AMD}^\text{c}$        & $\emptyset$                     & 0.3 $\text{AMD}^\text{s}$ & $\emptyset$                          & 0.3         & c2         & quick      \\
			Ac03-Ac03\_c2\_quick        & F3        & 0.3 $\text{AMD}^\text{c}$        & $\emptyset$                     & 0.3 $\text{AMD}^\text{c}$ & $\emptyset$                          & 0.3         & c2         & quick      \\
			Ac03-Ac03\_c4\_quick        & F4        & 0.3 $\text{AMD}^\text{c}$        & $\emptyset$                     & 0.3 $\text{AMD}^\text{c}$ & $\emptyset$                          & 0.3         & c4         & quick      \\
			Ac03-As03\_c4\_c2           & F5        & 0.3 $\text{AMD}^\text{c}$        & $\emptyset$                     & 0.3 $\text{AMD}^\text{s}$ & $\emptyset$                          & 0.3         & c4         & c2         \\
			Ac03-As03\_c4\_quick        & F6        & 0.3 $\text{AMD}^\text{c}$        & $\emptyset$                     & 0.3 $\text{AMD}^\text{s}$ & $\emptyset$                          & 0.3         & c4         & quick      \\
			Ac03-As03\_c2\_c2           & F7        & 0.3 $\text{AMD}^\text{c}$        & $\emptyset$                     & 0.3 $\text{AMD}^\text{s}$ & $\emptyset$                          & 0.3         & c2         & c2         \\
			\midrule
			\multicolumn{9}{l}{\textit{David et al. (2023)}}                                                                                                                                                                        \\
			nomodel\_c4\_c2             & NA        & $\emptyset$                      & $\emptyset$                     & $\emptyset$               & $\emptyset$                          & $\emptyset$ & c4         & c2         \\
			nomodel\_c2\_c2             & NB        & $\emptyset$                      & $\emptyset$                     & $\emptyset$               & $\emptyset$                          & $\emptyset$ & c2         & c2         \\
			A05+B06-A05+B06\_2L\_c2\_c2 & MA        & 0.6 AMD                          & 0.5 Bard                        & 0.6 AMD                   & 0.4 Bard                             & 0.15        & c2         & c2         \\
			Sc1-\_c4\_c2                & SA        & $\emptyset$                      & 1.0 $\text{Sim}^\text{c}$       & $\emptyset$               & 1.0 Sim                              & $\emptyset$ & c4         & c2         \\
			S1-\_c4\_c2                 & SB        & $\emptyset$                      & 1.0 $\text{Sim}^\text{c}$       & $\emptyset$               & $\emptyset$                          & $\emptyset$ & c4         & c2         \\
			\bottomrule
		\end{tabular}}
	\caption{T-LES models presented in this work. The $\cdot^c$ superscript denotes compressible formulations, $\cdot^s$ represents scalar models. "Bard" denotes the Bardina model, "Grad" the Gradient model, and "Sim" the scale similarity model.}
	\label{tab_models}
\end{table*}
\subsection{Effects of the mesh and numerical schemes}
\subsubsection{First-order statistics}

\begin{figure*}[htbp]
	\begin{center}
		\includegraphics[width=\textwidth]{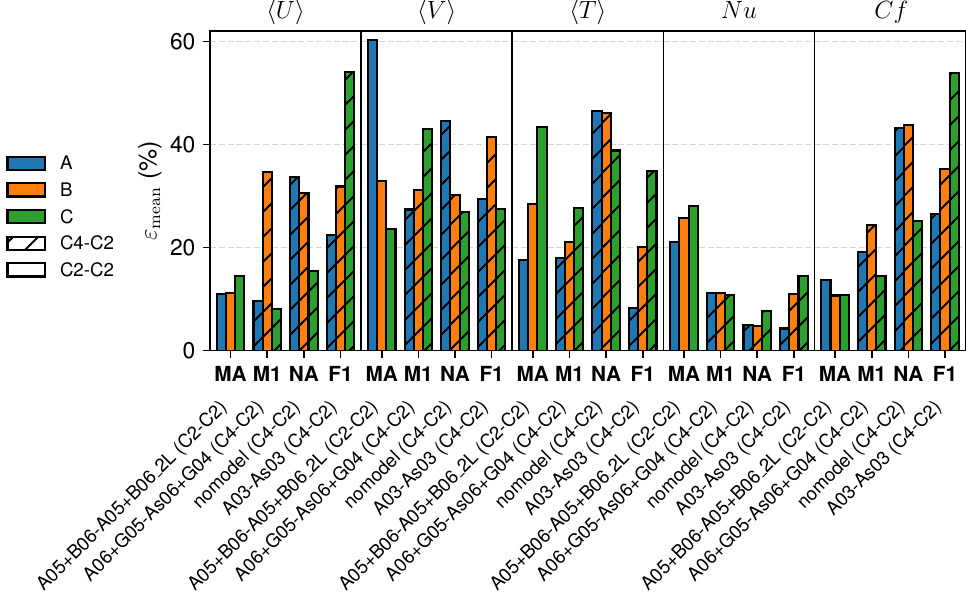}
	\end{center}
	\caption{Relative error on first-order statistics ($\langle U \rangle$, $\langle V \rangle$, $\langle T \rangle$, $Nu$, $C_f$) for the four selected models (MA, M1, NA, F1).}
	\label{fig_err_mean}
\end{figure*}

\begin{rewrite}
	In this section, we discuss the error rates across the mean statistics $\langle U \rangle$, $\langle V \rangle$, $\langle T \rangle$, $Nu$, and $C_f$. Figure~\ref{fig_err_mean} shows relative error bars for mean quantities $U, V, T, Nu, C_f$ for all selected models. The selected models exhibit acceptable accuracy on evaluated quantities. The finest mesh (A) exhibits the lowest error rate more often than coarser meshes (B) and (C). This is notable on temperature $T$ and streamwise velocity $U$, and is the consistent expected behavior for LES. Comparing mixed models MA and M1 exhibit similar performance on the velocity and temperature error. However, on wall quantities $Nu$ and $C_f$, they diverge from the rest of the selected models. The reference model MA achieves the lowest error on the friction coefficient (11.68\%) when compared to the M1 model (19.34\%). On the contrary, the M1 model is more accurate for the Nusselt number, exhibiting an error rate of 11.07\% against the MA model 24.94\%.
	The selected functional model F1 resolves the mean temperature $T$ best, and achieves the lowest overall error on this quantity (21.06\%), and performing best on the A mesh (8.24\%).
	Finally, the no-model simulation (NA) outperforms all other models on the Nusselt number $Nu$, with a mean error rate of 5.81\% (A: 4.9, B: 4.8, C: 7.7\%).
\end{rewrite}

\begin{figure*}[htbp]
	\begin{center}
		\includegraphics[width=1.0\textwidth]{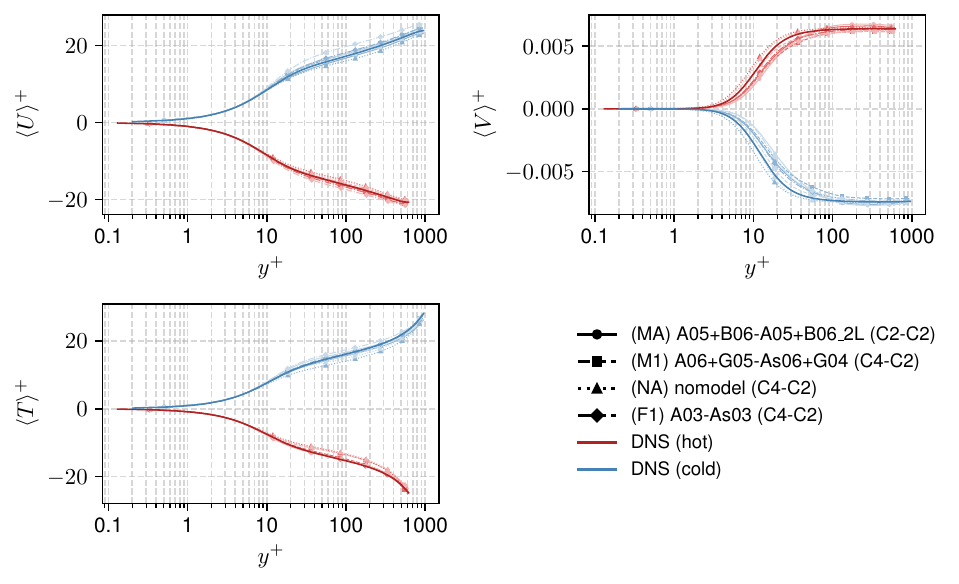}
	\end{center}
	\caption{Mean streamwise velocity \textbf{$\langle U \rangle^+$}, wall-normal velocity \textbf{$\langle V \rangle^+$}, and temperature \textbf{$\langle T \rangle^+$} profiles for the selected simulations. Coarser mesh resolutions are represented by lighter line shades.}
	\label{fig_u_v_t}
\end{figure*}

\begin{rewrite}

	Figure~\ref{fig_u_v_t} shows the mean streamwise velocity $\langle U \rangle^+$, wall-normal velocity $\langle V \rangle^+$, and temperature $\langle T \rangle^+$ profiles. For the streamwise velocity $\langle U \rangle^+$, all models exhibit good agreement with the DNS outside the viscous sublayer ($y^+ \geq 10$). As highlighted by~\citet{David2023TLES}, the no-model simulations (NA) underestimate the streamwise velocity profile on both sides. On the other hand, the functional model (F1) overestimates it. This is a consequence of the purely dissipative nature of functional models, as this indicates the computed friction velocity $u_\tau$ is lower than the DNS. Coarser grids, as shown by lighter shades, degrades the accuracy across all presented models on these quantities. The plotted mixed models (MA and M1) show good agreement with the DNS.

	Concerning the wall-normal velocity $\av{V}^+$, all simulations manages to capture the plateau around $y^+\geq 100$. The no-model simulation (NA) reproduces this behavior accurately across all meshes. The fine mesh (A) aligns closely with the DNS, while the coarse mesh (C) deviates. This is especially clear on the cold side, due to the higher local friction Reynolds number.
	All-normal velocity $\av{V}^+$, all simulations manage to capture the negative plateau at $y^+\geq 100$. The no-model (NA) simulation is able to capture the slope and plateau in close agreement with the DNS, even on the cold side. Conversely, the mixed models (MA) and (M1), and the functional model (F1) do not manage to accurately capture the transition region ($10 \leq y^+ \leq 100$), underestimating the wall-normal velocity dip before eventually recovering toward the plateau.

	The temperature $\langle T \rangle^+$ shows similar tendencies to the streamwise velocity. The no-model (NA) simulation is consistently below the DNS, while the functional model (F1) is above the reference.
	The mixed models are close to the reference, and the M1 model exhibits an improvement over the reference MA with a slightly more accurate temperature profile on both the hot and cold sides for the plotted meshes, especially on the cold side. On the hot side, the MA model performs slightly better than the presented M1 model.
	Overall, the mixed models provide a robust approach for the prediction of the mean profiles.

\end{rewrite}
\subsubsection{Second order statistics}

Second-order statistics pose a harder challenge to simulate.

\begin{figure*}[htbp]
	\begin{center}
		\includegraphics[width=1.0\textwidth]{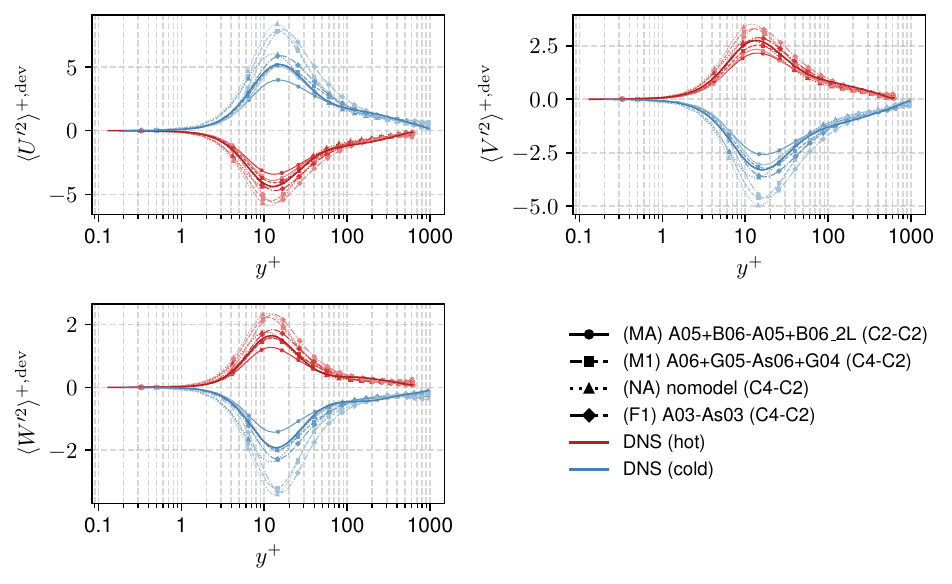}
	\end{center}
	\caption{Second-order statistics for the diagonal components of the deviatoric Reynolds stress tensor ($\langle U'^2 \rangle^{+,\mathrm{dev}}$, $\langle V'^2 \rangle^{+,\mathrm{dev}}$, $\langle W'^2 \rangle^{+,\mathrm{dev}}$)}
	\label{fig_rms_uvw}
\end{figure*}

\begin{rewrite}

	Figure~\ref{fig_rms_uvw} shows the wall-normal profiles of the deviatoric part of the Reynolds stress tensor.
	To further isolate the effect of each mesh, and closure model, the detailed profiles of the Reynolds stress tensors are provided in the appendix~\ref{sec:annexe_diag_part_rey_ten}.
	The DNS has a peak near $y^+\approx 15$, with the cold side reaching a higher magnitude than the hot side.

	For the streamwise stress $\av{U'^2}^{+,\rm{dev}}$, the cold side exhibits a dispersion between the different models. The functional model F1 and no-model NA closures over predict the streamwise velocity variance, with their peaks approaching 8, compared to the DNS peak at 5. On the other hand, the models on the hot side are less spread out, and better approximate this quantity. While the same trends are observable, the F1 and NA models over predict the peak value at -6 instead of -4.5.
	The MA model consistently under predicts the velocity variance peak on both the hot and cold sides, with the fine mesh performing worse than the coarse one.
	Conversely, the M1 model exhibits a better accuracy on both sides, even though the predicted peak shifts towards the outer layer slightly when compared to the DNS.
	The other diagonal parts of the Reynolds stress tensor $\av{V}^{+,\rm{dev}}$, and $\av{W}^{+,\rm{dev}}$ exhibit similar trends.

\end{rewrite}

\begin{figure*}[htbp]
	\begin{center}
		\includegraphics[width=1.0\textwidth]{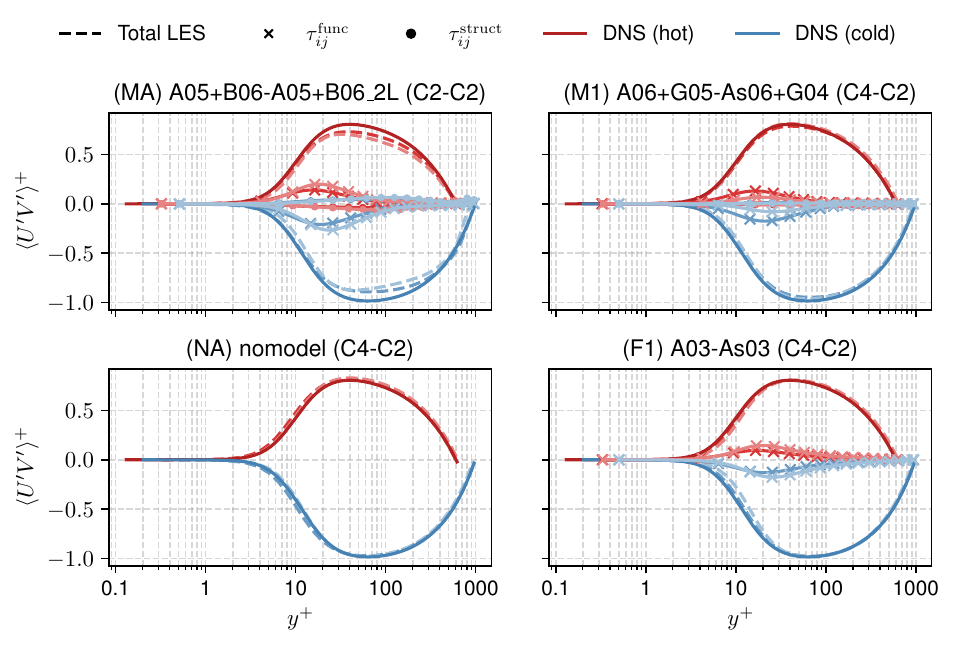}
	\end{center}
	\caption{Streamwise wall-normal Reynolds shear stress $\langle U'V' \rangle^+$ and corresponding subgrid-scale structural and functional closure terms for the selected models.}
	\label{fig_rms_uv}
\end{figure*}

\begin{rewrite}

	Figure~\ref{fig_rms_uv} shows the turbulent friction $\av{U'V'}$ for the 4 selected models, and the amplitude of the closure terms. The no-model simulation (NA) captures the behavior of this quantity accurately, with slight deviation on the cold side. This indicates that the simulation grids are fine enough to approximate this quantity without the need for a closure term. The MA model under predicts this quantity as the exit of the viscous sublayer $y^+\geq 7$, with the structural closure degrading the results, as it removes from $\av{U'V'}$. Conversely, the M1 and F1 closures both exhibit good agreement with the DNS, as the structural model does not interfere with the quantity in both cases. In the last two cases, the F1 and M1 models have a close to zero structural closure, enabling a better approximation of this quantity as opposed to the MA model. This indicates that the Gradient model performs better in this case.

\end{rewrite}

\begin{rewrite}

	For the streamwise convection term $\av{U'T'}$, the NA no-model and F1 functional closures exhibit sensitivity to the used mesh, particularly towards the cold wall. On that side, the coarsest mesh (C) over predicts the streamwise temperature transport. On the other hand, the hot side is more constrained with a lesser dispersion around the DNS data, although the same over prediction tendency is visible. The reference mixed model MA under predicts the peak temperature transport on both sides, with the coarse mesh exhibiting a better fit to the DNS.
	The M1 model exhibits a dispersion similar to the NA and F1 models, especially on the coarse grid (C), however, this model manages to stay close to the DNS reference, on the fine grid.

	The wall-normal turbulent convection $\av{V'T'}$ behaves similarly to the wall-normal velocity stress $\av{V'^2}$, exhibiting a plateau around $y^+\geq 100$. This time, the performance of the various models is closer to the one observed for $\av{V'^2}$, as on both the hot and cold sides, the (NA) no-model simulation captures best the transition between the viscous sublayer and the outer region of the flow $10\leq y^+ \leq 100$. Amongst the LES with closure models, the M1 model on the (C) mesh captures this quantity best, with small deviations from the reference. The finest mesh degrades the performance of this model. Models F1 and MA struggle to approximate this quantity when compared the M1 (C) mesh simulation and the no-model (NA) simulations.

	Finally, for the temperature variance $\av{T'^2}$, because there is no explicit subgrid-scale variance closure modeled in these simulations, the temperature variance serves as a pure diagnostic of how well the momentum and scalar closures are dynamically coupled. The errors seen here are the compounded result of the resolved velocity and temperature fields interacting. No-model and F1 simulations perform poorly, and are greatly affected by grid resolution in two ways.
	\rewriteFrancoise{\begin{enumerate}
			\item The amplitude of the near-wall peak is over predicted.
			\item In the logarithmic region, the temperature variance peak is over predicted as $y^+\rightarrow 1000$.
		\end{enumerate}}{The amplitude of the peak in both the near-wall and logarithmic region are over predicted.}
	This behavior is similar on both sides, and more severe on the cold side, and is worsened on a coarse mesh.
	Conversely, the MA model is slightly overdissipative on both sides, but this behavior helps the prediction in the log region.
	Finally, the M1 model performs the best to obtain the near-wall peak, and over estimates the end of the log-region behavior, with the coarse mesh faring poorly.

	\rewriteFrancoise{Additional plots are given in the appendix for}{Additional graphs are provided for} the temperature transport quantities $\av{U'T'}^+$, $\av{V'T'}^+$, and the temperature variance $\av{T'^2}^+$ in annexe~\ref{sec:annexe_temp_transport}.

\end{rewrite}

\begin{figure*}[htbp]
	\begin{center}
		\includegraphics[width=1.0\textwidth]{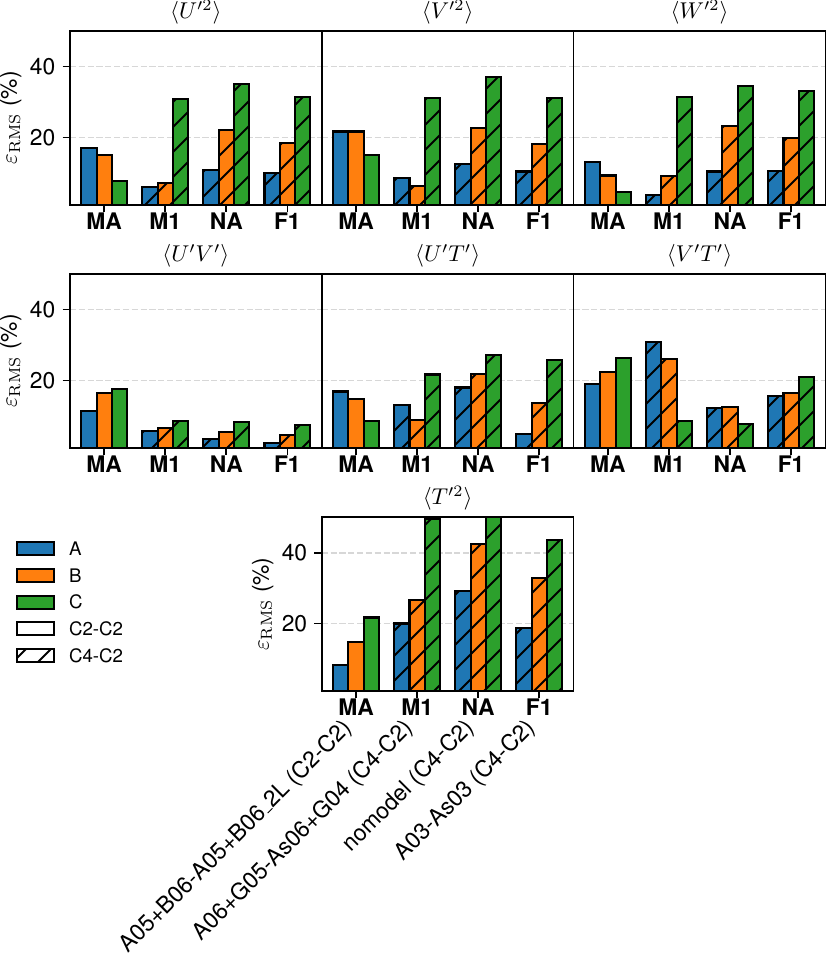}
	\end{center}
	\caption{Relative error on {second-order} statistics for the selected models (MA, M1, NA, F1).}
	\label{fig_err_rms}
\end{figure*}
\begin{figure*}[htbp]
	\begin{center}
		\includegraphics[width=1.0\textwidth]{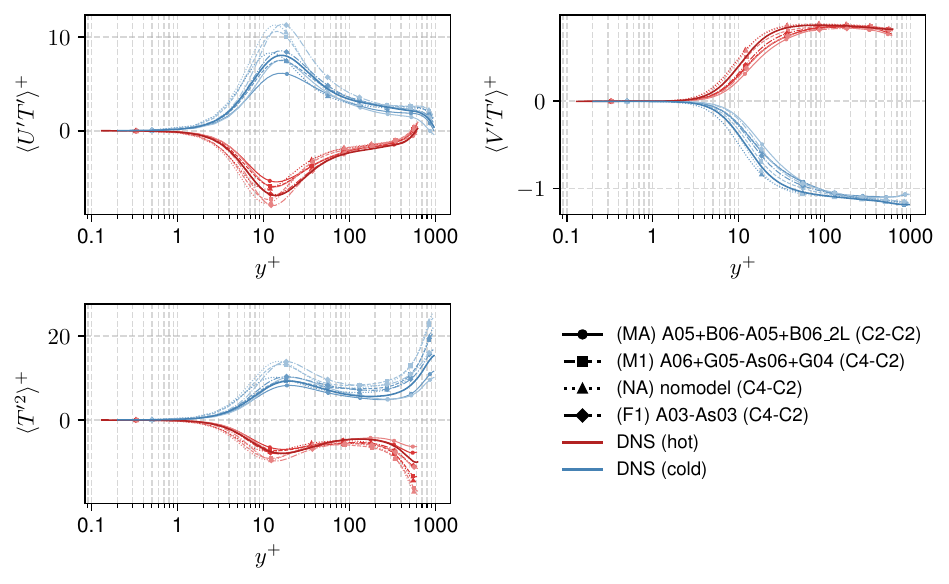}
	\end{center}
	\caption{Temperature transport terms ($\langle U'T' \rangle^+$, $\langle V'T' \rangle^+$) and temperature variance ($\langle T'^2 \rangle^+$), alongside their corresponding subgrid-scale closure terms, for the selected models.}
	\label{fig_rms_t}
\end{figure*}
\subsection{Unstable models}
\begin{rewrite}
	Many T-LES models exhibited numerical instability during testing.
	Table~\ref{tab_unstable_models} summarizes the unstable models encountered under both conditions.

	The Reynolds stress closure $\tau_{ij}$, the coupling of the AMD compressible and Bardina models was not numerically stable on the finest mesh (A). This suggests the grid resolution was too fine to provide sufficient numerical dissipation.
	Furthermore, increasing the Bardina closure constant from 0.4 to 0.6 destabilized the mixed models on meshes A and B.

\end{rewrite}
\begin{table*}
	\begin{center}
		\resizebox{\textwidth}{!}{
			\begin{tabular}{*{10}{c}}
				Model name                         & \multicolumn{2}{c}{$\tau$ model} & \multicolumn{2}{c}{$\pi$ model} & Functional Constant & \multicolumn{2}{c}{Numerical scheme} & Unstable meshes                                                \\
				$C\cdot$ Functional                &                                  & $C\cdot$Structural              & $C\cdot$Functional  & $C\cdot$Structural                   & Center          & Momentum convection & mass convection &      \\
				\toprule
				$\text{Ac06+B05-As06+B04\_c2\_c2}$ & 0.6 $\text{AMD}^{c}$             & 0.5 Bard                        & $\text{AMD}$        & 0.4 Bard                             & 0.15            & c2                  & c2              & A    \\
				$\text{Ac06+B05-As06+B04\_c4\_c2}$ & 0.6 $\text{AMD}^{c}$             & 0.5 Bard                        & $\text{AMD}$        & 0.4 Bard                             & 0.15            & c4                  & c2              & A    \\
				$\text{Ac06+B06-As06+B06\_c2\_c2}$ & 0.6 $\text{AMD}^{c}$             & 0.6 Bard                        & $\text{AMD}$        & 0.6 Bard                             & 0.15            & c2                  & c2              & A, B \\
				$\text{Ac06+B06-As06+B06\_c4\_c2}$ & 0.6 $\text{AMD}^{c}$             & 0.6 Bard                        & $\text{AMD}$        & 0.6 Bard                             & 0.15            & c4                  & c2              & A, B \\
				\bottomrule
			\end{tabular}}
	\end{center}
	\caption{Unstable models}
	\label{tab_unstable_models}
\end{table*}

\begin{rewrite}
	\section{Conclusion}
	\label{sec_conclusion}
	This work evaluates twelve new T-LES models (five mixed and seven functional) to assess their error rate on velocity and temperature statistics in an asymmetrically heated channel flow.
	Based on an established error rate function, the first and second-order statistics across three different grid resolutions, and four different numerical scheme combinations are assessed, highlighting the effects of each.

	For first-order statistics, mixed models provide the most accurate evaluations. However, performance diverges depending on the quantity.
	The reference mixed model used as the best known comparison in this setup (MA, using the AMD and Bardina closures) yields the lowest error for the skin friction coefficient $C_f$. Conversely, the proposed M1 model (utilizing the AMD-AMD Scalar and Gradient closures) improves the predictions of wall heat flux.

	Notably, the no-model baseline (NA) surprisingly achieves the lowest global error for the Nusselt number (Nu) across all tested meshes.

	Second-order statistics exhibit more noticeable difference with the no-model and functional closures.
	On the higher Reynolds number cold wall, the functional (F1) and no-model (NA) simulations are sensitive to grid resolution, exhibiting over predictions on the diagonal part of the Reynolds tensor $\av{U_j'^2}$, and temperature transport terms $\av{U_j'T'}$.
	The inclusion of structural closure models was important to dampen the effects of the overdissipative functional closures.

	The difference between the mixed models M1 and MA is akin to a trade-off, as the M1 model dampens the second-order statistics spikes (at $y^+\approx 10$), in effect decreasing the spikes higher in the channel.
	The proposed M1 model shows balanced results as it more accurately captures the peak amplitudes of the velocity variances and thermal transport without over-dampening them.
	Furthermore, for the turbulent shear stress $\av{U'V'}$, the Gradient model in the M1 closure prevents the degradation observed using the Bardina model with MA.

	Finally, numerical stability is an important constraint for T-LES. In configurations without heat sinks, increasing the structural model constant or using higher-order numerical schemes led to numerical instability. Specifically, the coupling of the compressible AMD and Bardina models proved unstable on the finest mesh, and raising the structural constant from 0.4 to 0.6 further destabilized the flow on intermediate grids.

	While none of the tested models are highly accurate on all presented quantities, the M1 mixed model, combining the Gradient and AMD closures exhibits satisfactory results and is slightly better than the previous best MA model.
\end{rewrite}

\section*{Acknowledgement}
This work was made possible by the GENCI allocations A0132A05099, A0152A14652 and SS012A15404 for HPC resources. The authors acknowledge the CEA STMF for the development of the TrioCFD computational software. This work was financed thanks to the ANR through the ANR-21-CE50-0031 project.

\section*{Data availability statement}
The data that support the findings of this study are available from the corresponding author upon reasonable request.

\newpage
\FloatBarrier
\bibliographystyle{aipauth4-1}
\bibliography{references}

\appendix
\onecolumn
\section{Error rate tables}
\subsection{First-order, second-order, and total error rate}
\label{sec:annexe_error_sec}

\begin{table*}[htbp]
	\centering
	\begin{tabular}{lllcccc}
		\toprule
		Model & Category   & Scheme   & Mesh A        & Mesh B        & Mesh C        & Average       \\
		\midrule
		M1    & Mixed      & C4-C2    & \textbf{18.8} & 21.6          & 37.2          & \textbf{25.9} \\
		M2    & Mixed      & C4-C2    & 25.8          & 22.8          & 33.6          & 27.4          \\
		M3    & Mixed      & C2-C2    & 25.4          & 29.2          & 42.2          & 32.2          \\
		M4    & Mixed      & C2-C2    & 31.7          & 36.7          & 29.7          & 32.7          \\
		M5    & Mixed      & C4-C2    & 24.8          & \textbf{21.3} & 56.5          & 34.2          \\
		F1    & Functional & C4-C2    & 20.3          & 31.2          & 33.1          & 28.2          \\
		F2    & Functional & C2-QUICK & 19.7          & 39.5          & 46.4          & 35.2          \\
		F3    & Functional & C2-QUICK & 29.7          & 37.6          & 44.1          & 37.1          \\
		F4    & Functional & C4-QUICK & 34.0          & 35.9          & 42.9          & 37.6          \\
		F5    & Functional & C4-C2    & 34.1          & 36.1          & 44.9          & 38.3          \\
		F6    & Functional & C4-QUICK & 34.9          & 42.5          & 49.0          & 42.2          \\
		F7    & Functional & C2-C2    & 20.1          & 71.2          & 48.0          & 46.4          \\
		NA    & No-Model   & C4-C2    & 29.7          & 28.5          & 32.4          & 30.2          \\
		NB    & No-Model   & C2-C2    & 31.2          & 38.7          & 44.9          & 38.3          \\
		MA    & Mixed      & C2-C2    & 39.1          & 25.9          & \textbf{18.7} & 27.9          \\
		SA    & Structural & C4-C2    & 25.0          & 29.6          & 35.5          & 30.0          \\
		SB    & Structural & C4-C2    & 41.0          & 46.0          & 50.8          & 46.0          \\
		\bottomrule
	\end{tabular}
	\caption{Relative Total Error (\%) across all tested models and mesh resolutions.}
	\label{tab:err_total}
\end{table*}


\begin{table*}[htbp]
	\centering
	\begin{tabular}{lllcccc}
		\toprule
		Model & Category   & Scheme   & Mesh A        & Mesh B        & Mesh C        & Average       \\
		\midrule
		M1    & Mixed      & C4-C2    & 29.5          & 34.9          & 45.1          & 36.5          \\
		M2    & Mixed      & C4-C2    & 42.0          & 38.5          & 47.2          & 42.6          \\
		M3    & Mixed      & C2-C2    & 37.5          & 41.5          & 31.7          & 36.9          \\
		M4    & Mixed      & C2-C2    & 46.2          & 39.4          & 39.2          & 41.6          \\
		M5    & Mixed      & C4-C2    & 33.9          & 34.0          & 100.0         & 56.0          \\
		F1    & Functional & C4-C2    & 32.3          & 45.5          & 33.5          & 37.1          \\
		F2    & Functional & C2-QUICK & 31.3          & 68.1          & 53.7          & 51.0          \\
		F3    & Functional & C2-QUICK & 52.7          & 62.8          & 49.5          & 55.0          \\
		F4    & Functional & C4-QUICK & 57.8          & 55.9          & 49.4          & 54.3          \\
		F5    & Functional & C4-C2    & 59.9          & 60.7          & 54.7          & 58.4          \\
		F6    & Functional & C4-QUICK & 59.4          & 70.0          & 62.2          & 63.9          \\
		F7    & Functional & C2-C2    & 33.5          & 36.6          & 31.9          & 34.0          \\
		NA    & No-Model   & C4-C2    & 49.1          & 34.3          & 29.5          & 37.6          \\
		NB    & No-Model   & C2-C2    & 43.2          & 41.1          & 37.4          & 40.6          \\
		MA    & Mixed      & C2-C2    & 63.2          & 35.5          & \textbf{26.2} & 41.6          \\
		SA    & Structural & C4-C2    & 38.7          & 36.9          & 36.5          & 37.4          \\
		SB    & Structural & C4-C2    & \textbf{28.5} & \textbf{30.8} & 35.6          & \textbf{31.7} \\
		\bottomrule
	\end{tabular}
	\caption{Relative Error (\%) on first-order statistics across all tested models.}
	\label{tab:err_mean}
\end{table*}


\begin{table*}[htbp]
	\centering
	\begin{tabular}{lllcccc}
		\toprule
		Model & Category   & Scheme   & Mesh A       & Mesh B       & Mesh C        & Average       \\
		\midrule
		M1    & Mixed      & C4-C2    & 9.8          & 10.6         & 30.6          & 17.0          \\
		M2    & Mixed      & C4-C2    & 12.3         & \textbf{9.7} & 22.3          & \textbf{14.8} \\
		M3    & Mixed      & C2-C2    & 15.3         & 18.9         & 50.9          & 28.4          \\
		M4    & Mixed      & C2-C2    & 19.6         & 34.4         & 21.8          & 25.3          \\
		M5    & Mixed      & C4-C2    & 17.2         & 10.8         & 20.2          & 16.1          \\
		F1    & Functional & C4-C2    & 10.3         & 19.3         & 32.8          & 20.8          \\
		F2    & Functional & C2-QUICK & 10.0         & 15.6         & 40.3          & 22.0          \\
		F3    & Functional & C2-QUICK & 10.5         & 16.6         & 39.7          & 22.3          \\
		F4    & Functional & C4-QUICK & 14.2         & 19.2         & 37.5          & 23.6          \\
		F5    & Functional & C4-C2    & 12.7         & 15.6         & 36.7          & 21.6          \\
		F6    & Functional & C4-QUICK & 14.5         & 19.7         & 38.0          & 24.1          \\
		F7    & Functional & C2-C2    & \textbf{8.9} & 100.0        & 61.4          & 56.8          \\
		NA    & No-Model   & C4-C2    & 13.5         & 23.7         & 34.8          & 24.0          \\
		NB    & No-Model   & C2-C2    & 21.3         & 36.7         & 51.2          & 36.4          \\
		MA    & Mixed      & C2-C2    & 19.1         & 17.9         & \textbf{12.3} & 16.4          \\
		SA    & Structural & C4-C2    & 13.5         & 23.6         & 34.7          & 23.9          \\
		SB    & Structural & C4-C2    & 51.4         & 58.7         & 63.5          & 57.9          \\
		\bottomrule
	\end{tabular}
	\caption{Relative Error (\%) on second-order statistics across all tested models.}
	\label{tab:err_rms}
\end{table*}


\subsection{Mean error}
For each quantity in $U, V, T, Nu, C_f$, one table is given for the per-mesh error, and the average error across meshes.
\begin{table*}[htbp]
	\centering
	\begin{tabular}{lllcccc}
		\toprule
		Label & Category   & Scheme & Mesh A       & Mesh B        & Mesh C       & Average       \\
		\midrule
		MA    & Mixed      & C2-C2  & 11.0         & \textbf{11.2} & 14.5         & \textbf{12.3} \\
		M1    & Mixed      & C4-C2  & \textbf{9.6} & 34.7          & \textbf{8.1} & 17.5          \\
		NA    & No-Model   & C4-C2  & 33.7         & 30.5          & 15.4         & 26.5          \\
		F1    & Functional & C4-C2  & 22.4         & 31.8          & 54.1         & 36.1          \\
		\bottomrule
	\end{tabular}
	\caption{Relative error (\%) for the mean quantity U across the selected models.}
	\label{tab:err_selected_U}
\end{table*}


\begin{table*}[htbp]
	\centering
	\begin{tabular}{lllcccc}
		\toprule
		Label & Category   & Scheme & Mesh A        & Mesh B        & Mesh C        & Average       \\
		\midrule
		F1    & Functional & C4-C2  & 29.5          & 41.4          & 27.5          & \textbf{32.8} \\
		M1    & Mixed      & C4-C2  & \textbf{27.4} & 31.1          & 43.0          & 33.8          \\
		NA    & No-Model   & C4-C2  & 44.6          & \textbf{30.2} & 26.9          & 33.9          \\
		MA    & Mixed      & C2-C2  & 60.3          & 32.9          & \textbf{23.5} & 38.9          \\
		\bottomrule
	\end{tabular}
	\caption{Relative error (\%) for the mean quantity V across the selected models.}
	\label{tab:err_selected_V}
\end{table*}


\begin{table*}[htbp]
	\centering
	\begin{tabular}{lllcccc}
		\toprule
		Label & Category   & Scheme & Mesh A       & Mesh B        & Mesh C        & Average       \\
		\midrule
		F1    & Functional & C4-C2  & \textbf{8.2} & \textbf{20.1} & 34.8          & \textbf{21.1} \\
		M1    & Mixed      & C4-C2  & 17.9         & 21.0          & \textbf{27.7} & 22.2          \\
		MA    & Mixed      & C2-C2  & 17.7         & 28.4          & 43.3          & 29.8          \\
		NA    & No-Model   & C4-C2  & 46.6         & 46.2          & 38.8          & 43.9          \\
		\bottomrule
	\end{tabular}
	\caption{Relative error (\%) for the mean quantity T across the selected models.}
	\label{tab:err_selected_T}
\end{table*}


\begin{table*}[htbp]
	\centering
	\begin{tabular}{lllcccc}
		\toprule
		Label & Category   & Scheme & Mesh A       & Mesh B       & Mesh C       & Average      \\
		\midrule
		NA    & No-Model   & C4-C2  & 4.9          & \textbf{4.8} & \textbf{7.7} & \textbf{5.8} \\
		F1    & Functional & C4-C2  & \textbf{4.3} & 11.0         & 14.5         & 9.9          \\
		M1    & Mixed      & C4-C2  & 11.2         & 11.2         & 10.8         & 11.1         \\
		MA    & Mixed      & C2-C2  & 21.1         & 25.7         & 28.1         & 24.9         \\
		\bottomrule
	\end{tabular}
	\caption{Relative error (\%) for the mean quantity $Nu$ across the selected models.}
	\label{tab:err_selected_Nu}
\end{table*}


\begin{table*}[htbp]
	\centering
	\begin{tabular}{lllcccc}
		\toprule
		Label & Category   & Scheme & Mesh A        & Mesh B        & Mesh C        & Average       \\
		\midrule
		MA    & Mixed      & C2-C2  & \textbf{13.6} & \textbf{10.7} & \textbf{10.8} & \textbf{11.7} \\
		M1    & Mixed      & C4-C2  & 19.1          & 24.4          & 14.6          & 19.3          \\
		NA    & No-Model   & C4-C2  & 43.3          & 43.7          & 25.2          & 37.4          \\
		F1    & Functional & C4-C2  & 26.6          & 35.2          & 53.9          & 38.6          \\
		\bottomrule
	\end{tabular}
	\caption{Relative error (\%) for the mean quantity $C_f$ across the selected models.}
	\label{tab:err_selected_Cf}
\end{table*}

\subsection{Second-order error rate}
Similarly to the previous section, for each quantity in $U'^2, V'^2, W'^2, U'V', U'T', V'T', T'^2$, the error rate is given on each mesh and as an average of the mesh error.
\begin{table*}[htbp]
	\centering
	\begin{tabular}{lllcccc}
		\toprule
		Label & Category   & Scheme & Mesh A       & Mesh B       & Mesh C       & Average       \\
		\midrule
		MA    & Mixed      & C2-C2  & 17.0         & 15.0         & \textbf{7.6} & \textbf{13.2} \\
		M1    & Mixed      & C4-C2  & \textbf{6.0} & \textbf{7.0} & 30.8         & 14.6          \\
		NA    & No-Model   & C4-C2  & 10.8         & 22.1         & 35.1         & 22.7          \\
		F1    & Functional & C4-C2  & 9.9          & 18.4         & 31.5         & 19.9          \\
		\bottomrule
	\end{tabular}
	\caption{Relative error (\%) for the second-order quantity $\langle U'^2 \rangle$ across the selected models.}
	\label{tab:err_selected_urms}
\end{table*}


\begin{table*}[htbp]
	\centering
	\begin{tabular}{lllcccc}
		\toprule
		Label & Category   & Scheme & Mesh A       & Mesh B       & Mesh C        & Average       \\
		\midrule
		MA    & Mixed      & C2-C2  & 21.7         & 21.7         & \textbf{15.1} & 19.5          \\
		M1    & Mixed      & C4-C2  & \textbf{8.6} & \textbf{6.2} & 31.1          & \textbf{15.3} \\
		NA    & No-Model   & C4-C2  & 12.6         & 22.6         & 37.1          & 24.1          \\
		F1    & Functional & C4-C2  & 10.4         & 18.1         & 31.1          & 19.8          \\
		\bottomrule
	\end{tabular}
	\caption{Relative error (\%) for the second-order quantity $\langle V'^2 \rangle$ across the selected models.}
	\label{tab:err_selected_vrms}
\end{table*}


\begin{table*}[htbp]
	\centering
	\begin{tabular}{lllcccc}
		\toprule
		Label & Category   & Scheme & Mesh A       & Mesh B       & Mesh C       & Average      \\
		\midrule
		MA    & Mixed      & C2-C2  & 13.0         & 9.2          & \textbf{4.5} & \textbf{8.9} \\
		M1    & Mixed      & C4-C2  & \textbf{3.8} & \textbf{9.1} & 31.3         & 14.8         \\
		NA    & No-Model   & C4-C2  & 10.4         & 23.2         & 34.4         & 22.6         \\
		F1    & Functional & C4-C2  & 10.4         & 19.8         & 33.1         & 21.1         \\
		\bottomrule
	\end{tabular}
	\caption{Relative error (\%) for the second-order quantity $\langle W'^2 \rangle$ across the selected models.}
	\label{tab:err_selected_wrms}
\end{table*}


\begin{table*}[htbp]
	\centering
	\begin{tabular}{lllcccc}
		\toprule
		Label & Category   & Scheme & Mesh A       & Mesh B       & Mesh C       & Average      \\
		\midrule
		MA    & Mixed      & C2-C2  & 11.4         & 16.4         & 17.6         & 15.1         \\
		M1    & Mixed      & C4-C2  & 5.7          & 6.7          & 8.7          & 7.0          \\
		NA    & No-Model   & C4-C2  & 3.5          & 5.6          & 8.2          & 5.8          \\
		F1    & Functional & C4-C2  & \textbf{2.3} & \textbf{4.7} & \textbf{7.5} & \textbf{4.8} \\
		\bottomrule
	\end{tabular}
	\caption{Relative error (\%) for the second-order quantity $\langle U'V' \rangle$ across the selected models.}
	\label{tab:err_selected_uv}
\end{table*}


\begin{table*}[htbp]
	\centering
	\begin{tabular}{lllcccc}
		\toprule
		Label & Category   & Scheme & Mesh A       & Mesh B       & Mesh C       & Average       \\
		\midrule
		MA    & Mixed      & C2-C2  & 16.9         & 14.9         & \textbf{8.6} & \textbf{13.5} \\
		M1    & Mixed      & C4-C2  & 13.0         & \textbf{9.0} & 21.7         & 14.6          \\
		NA    & No-Model   & C4-C2  & 18.0         & 21.7         & 27.2         & 22.3          \\
		F1    & Functional & C4-C2  & \textbf{4.8} & 13.6         & 25.8         & 14.7          \\
		\bottomrule
	\end{tabular}
	\caption{Relative error (\%) for the second-order quantity $\langle U'T' \rangle$ across the selected models.}
	\label{tab:err_selected_u_theta}
\end{table*}


\begin{table*}[htbp]
	\centering
	\begin{tabular}{lllcccc}
		\toprule
		Label & Category   & Scheme & Mesh A        & Mesh B        & Mesh C       & Average       \\
		\midrule
		MA    & Mixed      & C2-C2  & 19.1          & 22.4          & 26.3         & 22.6          \\
		M1    & Mixed      & C4-C2  & 31.0          & 26.1          & 8.6          & 21.9          \\
		NA    & No-Model   & C4-C2  & \textbf{12.3} & \textbf{12.6} & \textbf{7.8} & \textbf{10.9} \\
		F1    & Functional & C4-C2  & 15.5          & 16.4          & 21.0         & 17.6          \\
		\bottomrule
	\end{tabular}
	\caption{Relative error (\%) for the second-order quantity $\langle V'T' \rangle$ across the selected models.}
	\label{tab:err_selected_v_theta}
\end{table*}


\begin{table*}[htbp]
	\centering
	\begin{tabular}{lllcccc}
		\toprule
		Label & Category   & Scheme & Mesh A       & Mesh B        & Mesh C        & Average       \\
		\midrule
		MA    & Mixed      & C2-C2  & \textbf{8.4} & \textbf{14.7} & \textbf{21.7} & \textbf{14.9} \\
		M1    & Mixed      & C4-C2  & 20.0         & 26.6          & 49.4          & 32.0          \\
		NA    & No-Model   & C4-C2  & 29.1         & 42.6          & 52.5          & 41.4          \\
		F1    & Functional & C4-C2  & 18.7         & 32.9          & 43.6          & 31.7          \\
		\bottomrule
	\end{tabular}
	\caption{Relative error (\%) for the second-order quantity $\langle T'^2 \rangle$ across the selected models.}
	\label{tab:err_selected_theta_rms}
\end{table*}

\section{Second order quantities with their closures}
\label{sec:annexe_diag_part_rey_ten}
\subsection{Diagonal part of the Reynolds stress tensor}
\newcommand{\closurefig}[4]{
	\begin{figure*}[htbp]
		\centering
		\begin{subfigure}[b]{\textwidth}
			\centering
			\includegraphics[width=\textwidth]{closures_#1_mesh_A.pdf}
			\caption{Mesh A (Fine)}
			\label{fig:#4_A}
		\end{subfigure}
		\begin{subfigure}[b]{\textwidth}
			\centering
			\includegraphics[width=\textwidth]{closures_#1_mesh_C.pdf}
			\caption{Mesh C (Coarse)}
			\label{fig:#4_C}
		\end{subfigure}
		\caption{Wall-normal profiles of #2 alongside the structural and functional closure terms across the selected models. Comparing the fine mesh (top) and coarse mesh (bottom) highlights the grid sensitivity of the #3 predictions.}
		\label{fig:#4}
	\end{figure*}
}

This section presents the detailed wall-normal profiles of the second-order statistics, explicitly plotting the contributions of the structural and functional closure terms for meshes A and C.


\closurefig{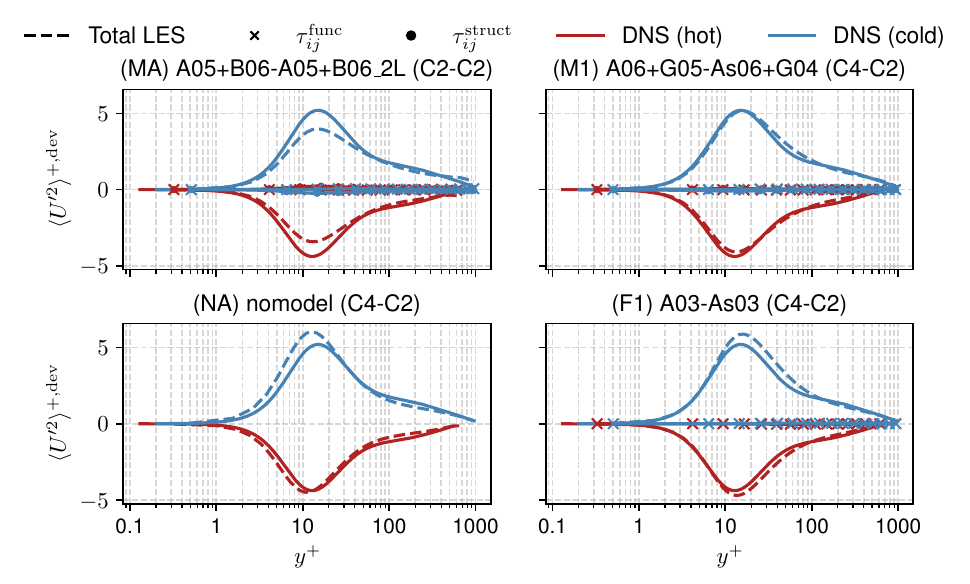}{$\langle U'^2 \rangle^{+,\mathrm{dev}}$}{streamwise variance}{closures_urms}

\closurefig{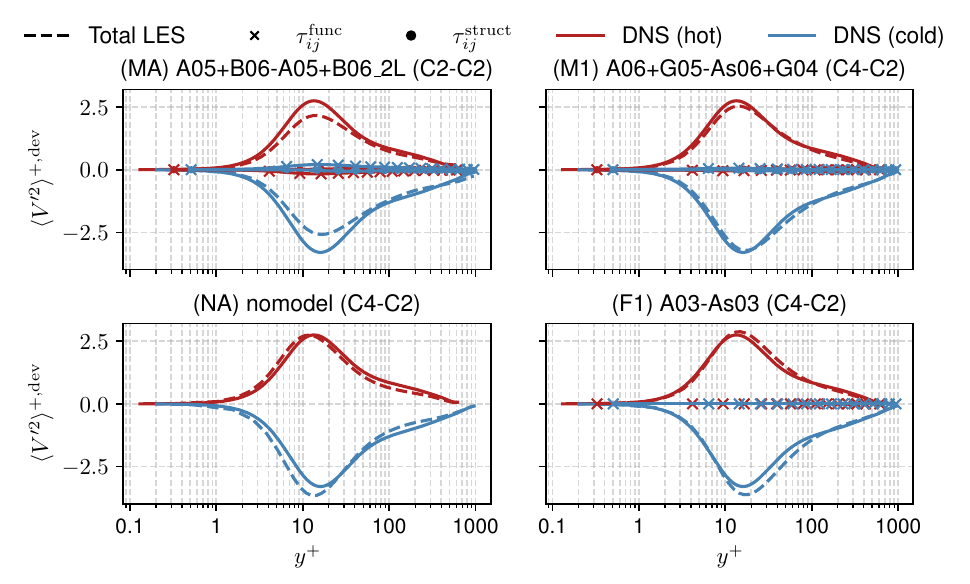}{$\langle V'^2 \rangle^{+,\mathrm{dev}}$}{wall-normal variance}{closures_vrms}

\closurefig{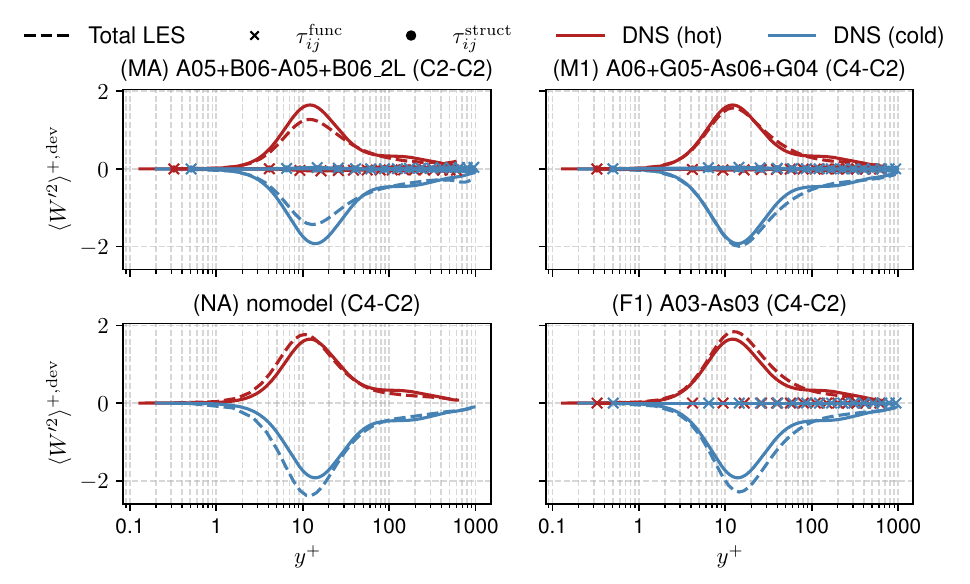}{$\langle W'^2 \rangle^{+,\mathrm{dev}}$}{spanwise variance}{closures_wrms}

\subsection{Temperature transport quantities and temperature variation}
\label{sec:annexe_temp_transport}

\closurefig{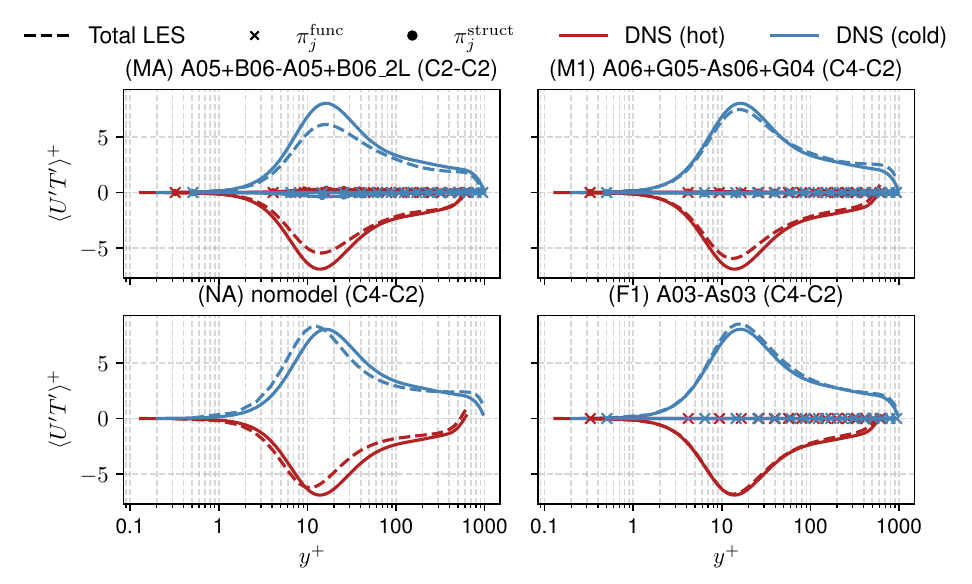}{$\langle U'T' \rangle^+$}{streamwise turbulent heat flux}{closures_utheta}

\closurefig{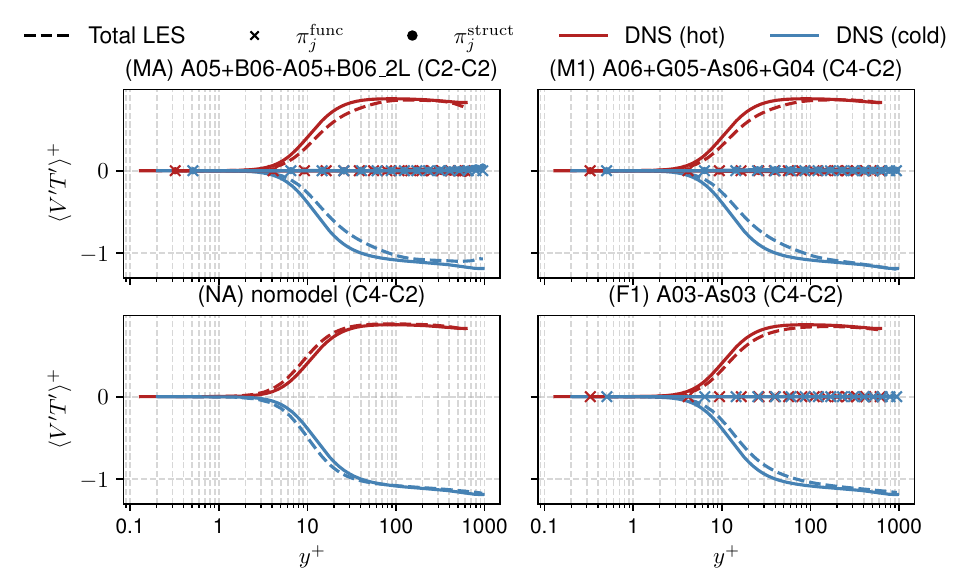}{$\langle V'T' \rangle^+$}{wall-normal turbulent heat flux}{closures_vtheta}

\closurefig{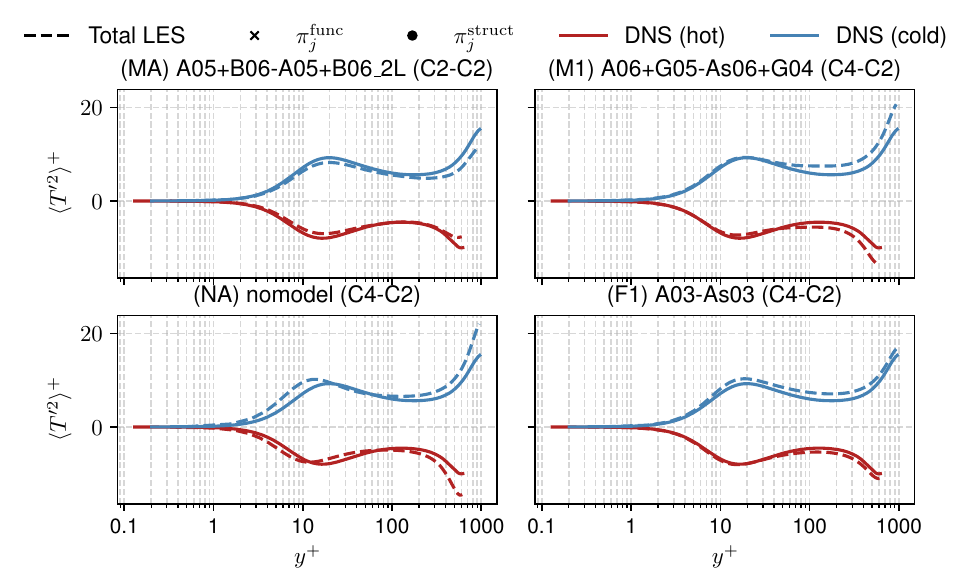}{$\langle T'^2 \rangle^+$}{temperature variance}{closures_thetatarms}

\end{document}